\documentclass{article}

\usepackage{arxiv}
\usepackage{upquote}

\usepackage{graphicx}
\usepackage[utf8]{inputenc} 
\usepackage[T1]{fontenc}    
\usepackage{hyperref}       
\usepackage{url}            
\usepackage{booktabs}       
\usepackage{amsfonts}       
\usepackage{nicefrac}       
\usepackage{microtype}      
\usepackage{ulem}
\usepackage{listings}
\usepackage{xcolor}

\definecolor{codegreen}{rgb}{0,0.6,0}
\definecolor{codegray}{rgb}{0.5,0.5,0.5}
\definecolor{codepurple}{rgb}{0.58,0,0.82}
\definecolor{backcolour}{rgb}{0.95,0.95,0.92}

\lstdefinestyle{mystyle}{
    backgroundcolor=\color{backcolour},   
    commentstyle=\color{codegreen},
    keywordstyle=\color{magenta},
    numberstyle=\tiny\color{codegray},
    stringstyle=\color{codepurple},
    basicstyle=\ttfamily\footnotesize,
    breakatwhitespace=false,         
    breaklines=true,                 
    captionpos=b,                    
    keepspaces=true,                 
    numbers=none,                    
    numbersep=5pt,                  
    showspaces=false,                
    showstringspaces=false,
    showtabs=false,                  
    tabsize=2
}

\title{Adding Uncertainty to Neural Network Regression Tasks in the Geosciences}

\author{%
  Elizabeth A. Barnes\thanks{Corresponding author webpage: http://barnes.atmos.colostate.edu} \\
  Department of Atmospheric Science\\
  Colorado State University\\
  Fort Collins, CO 80526 \\
  \texttt{eabarnes@colostate.edu} \\
  \And
  Randal J. Barnes \\
  Civil, Environmental, and Geo- Engineering\\
  University of Minnesota\\
  Minneapolis, MN, USA\\
  \texttt{barne003@umn.edu}
  \And
  Nicolas Gordillo \\
  Department of Atmospheric Science\\
  Colorado State University\\
  Fort Collins, CO 80526 \\
  \texttt{nicojg@rams.colostate.edu} \\
}

\begin{document}
\maketitle

\begin{abstract} 
A simple method for adding uncertainty to neural network regression tasks via estimation of a general probability distribution is described. The methodology supports estimation of heteroscedastic, asymmetric uncertainties by a simple modification of the network output and loss function. Method performance is demonstrated with a simple one dimensional data set and then applied to a more complex regression task using synthetic climate data. 
\end{abstract}
\clearpage

\section{Motivation}
Uncertainty quantification via machine learning methods is a vibrant area of research, and many methods for uncertainty estimation via artificial neural networks have been proposed \cite{Abdar2020}. To date, the majority of studies have focused on uncertainty quantification for classification. However, problems in the geosciences are often framed in terms of regression tasks (i.e. estimating continuous values). For example, geoscientists might wish to estimate how much it will rain tomorrow, the peak discharge of a river next year, or the thickness of an aquifer at a set of locations.  Estimates of the uncertainty associated with the predicted values are necessary to formulate policy, make decisions, and assess risk.

Here, we describe and demonstrate a simple method for adding uncertainty to most any neural network regression architecture. The method works by tasking the network to locally predict the parameters of a user-specified probability distribution, rather than just predict the value alone. Although the approach is a standard in the computer science literature \cite{Duerr2020}, it is much less known in the geoscience community. A few exceptions include \cite{Foster2021}, \cite{BarnesBarnes2021} and \cite{Guillaumin2021}, who present discussions of this approach with additional complexities due to their specific applications. This method for incorporating uncertainty is understandable, surprisingly general, and simple to implement with confidence. We believe that this method will become a powerful go-to approach moving forward.

\section{Background}\label{S_back}
In this note we attempt to explain and implement {\it multivariate nonlinear heteroscedastic parametric regression} using a deep neural network \cite{Duerr2020,Nix1994,Nix1995,Williams1996}. That many syllables is far too big a mouthful to swallow in one go so, like eating an elephant\footnote{To borrow a quote by Desmond Tutu.}, we are going to tackle this one bite at a time. 

We will demonstrate the concepts using synthetic data sets. The code fragments used to generate these synthetic data sets are given in Appendix~\ref{A_code}.

\subsection{Regression}
\begin{figure}[!htb]
  \centering
  \includegraphics[width=3.5in]{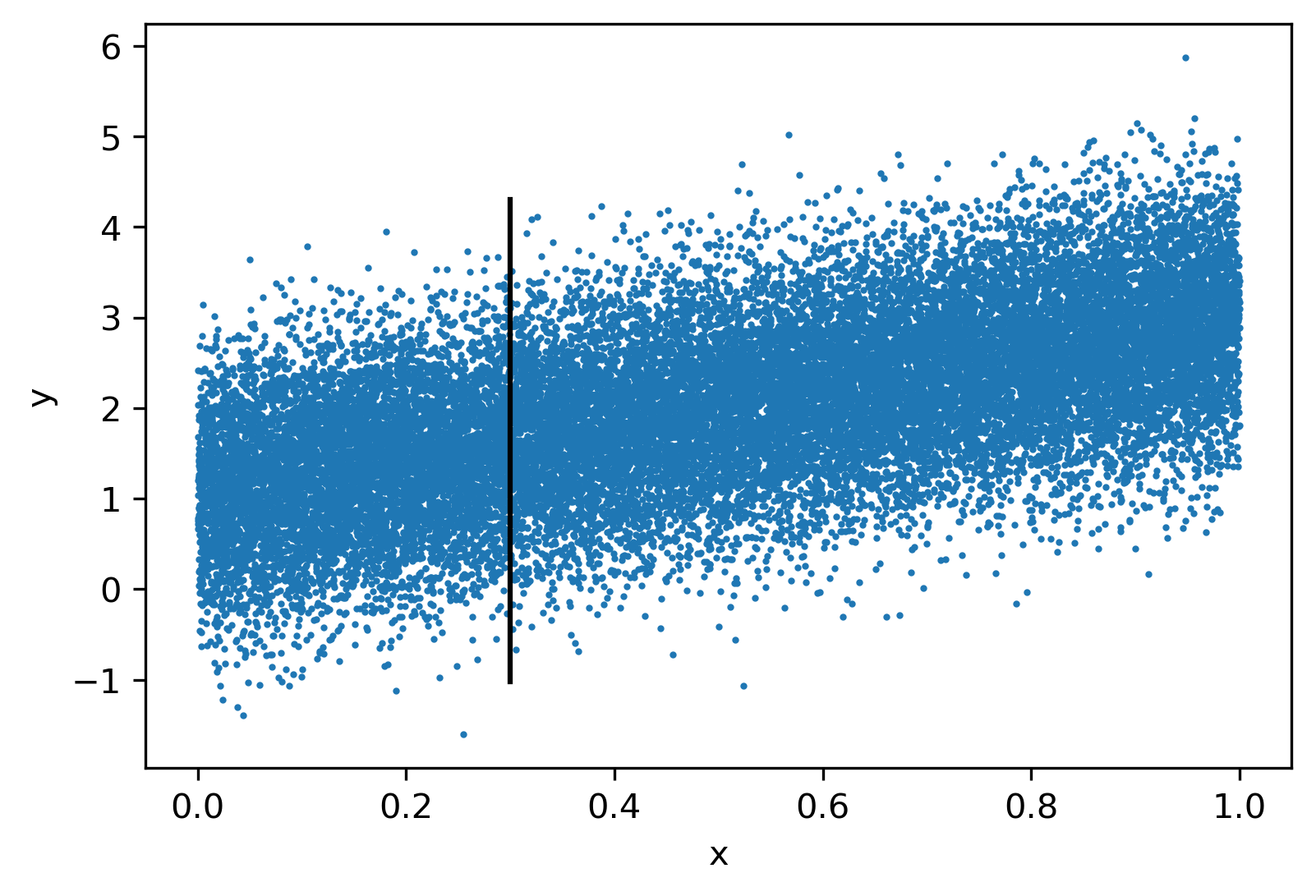}
  \caption{A synthetic data set containing $25,000$ sample $(x, y)$ pairs. The vertical black line indicates the location of $x = 0.3$. A rising trend is clearly visible: on average, as $x$ increases $y$ increases.}
  \label{bg_01}
\end{figure}

Consider the synthetic data set shown in Figure~\ref{bg_01}, which contains $25,000$ sample $(x, y)$ pairs.  The cloud of data is so dense that only a small fraction of the samples on the outer edges are distinguishable as individual points. Nonetheless, a rising trend is clearly visible: on average, as $x$ increases $y$ increases.

The usual stated goal of {\it regression} is to predict $y$ as a function of $x$.  Consider $y$ at $x = 0.3$, which is indicated by the vertical black line in Figure~\ref{bg_01}. Around $x = 0.3$ there is a wide distribution of $y$ values. Which value of $y$ are we attempting to predict at $x = 0.3$? The mean?  The median?

In this note, our goal for {\it regression} is not to predict $y$, but to predict the conditional distribution of $y$ as a function of $x$. With the conditional distribution we can compute the associated conditional mean, the conditional median, the conditional $90$'th percentile, or the ... as needed for the current question at hand.

\subsection{Parametric}
With {\it parametric} regression, we select the form of the conditional distribution before training. This is a modeling choice. The parameters defining the distribution as a function of $x$ are determined during training. As discussed in Section~\ref{S_loss}, we will use the form of the conditional distribution when defining our loss function.

\begin{figure}[!htb]
  \centering
  \includegraphics[width=3.5in]{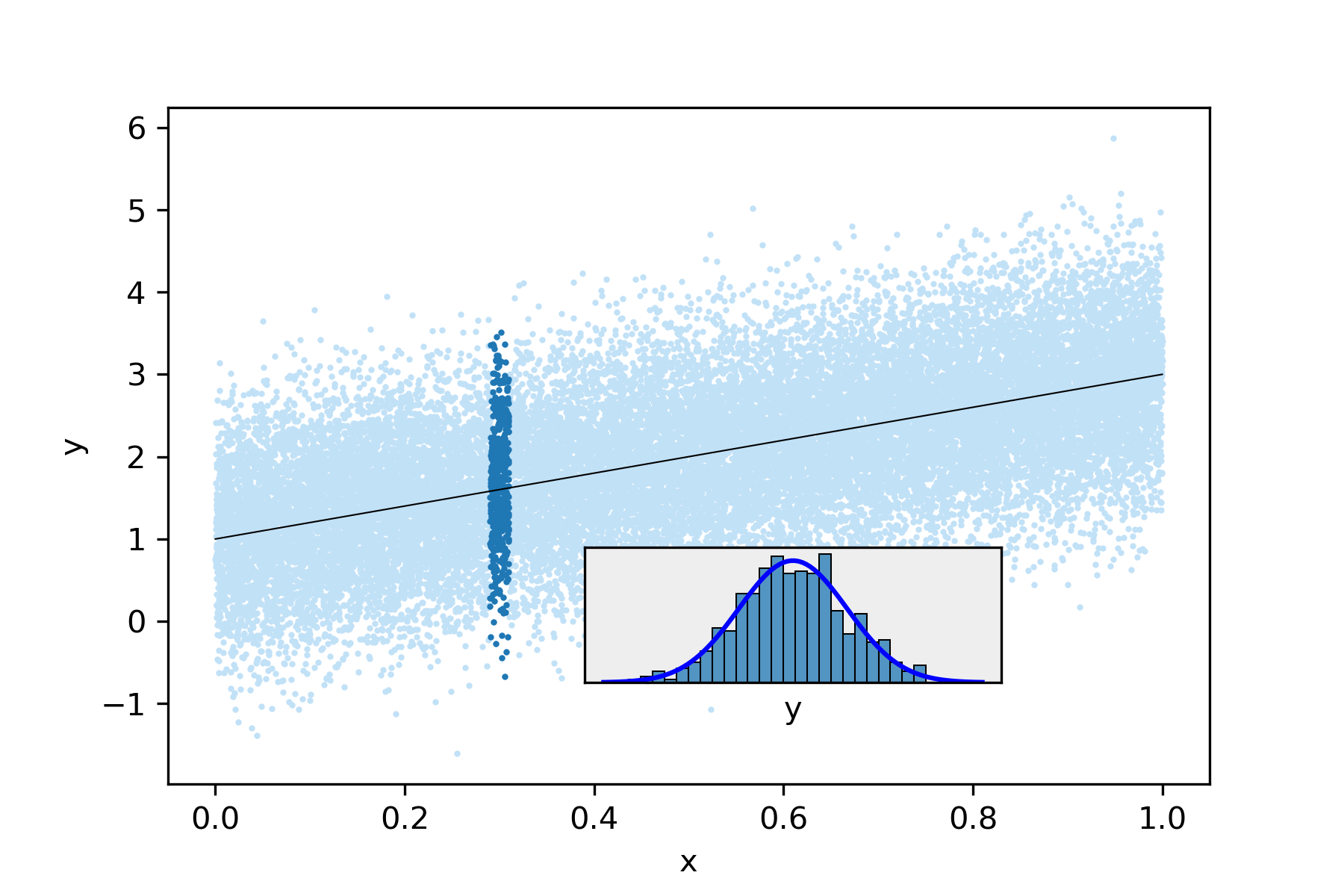}
  \caption{A synthetic data set containing $25,000$ sample $(x, y)$ pairs. These are the same data given shown Figure~\ref{bg_01}. The data in the subset, range $0.29 < x < 0.31$, are highlighted. The inset shows the histogram for the $y$ values in this subset along with the computed best-fit normal probability density function. The thin black line shows the conditional expected value of $y$ given $x$.}
  \label{bg_02}
\end{figure}

The most common choice is the normal distribution, where the mean and standard deviation are the defining parameters. Figure~\ref{bg_02} displays the same data given in Figure 1. The data subset in the range $0.29< x < 0.31$ are highlighted. The inset shows the histogram for the $y$ values in this subset along with the computed best-fit normal probability density function.

In almost all applications of curve-fitting software, the conditional distributions of the $y$ values are modeled using the normal distribution. When we use a simple least squares (or root-mean-square error) loss function, we implicitly model the conditional distributions of the $y$ values as a normal distribution\footnote{Minimizing the sum of the squared residuals is optimal if the conditional distribution is normal and homoscedastic.  See, for example, \cite[Section 4.3]{Duerr2020}.}. However, the normal distribution is not always the best choice. 

The normal distribution is symmetric about the mean and its tail extends below $0$.  As such, a normal distribution may not be appropriate when we model strictly positive quantities like length, mass, concentration, and rates (e.g. precipitation).  

\begin{figure}[!htb]
  \centering
  \includegraphics[width=3.5in]{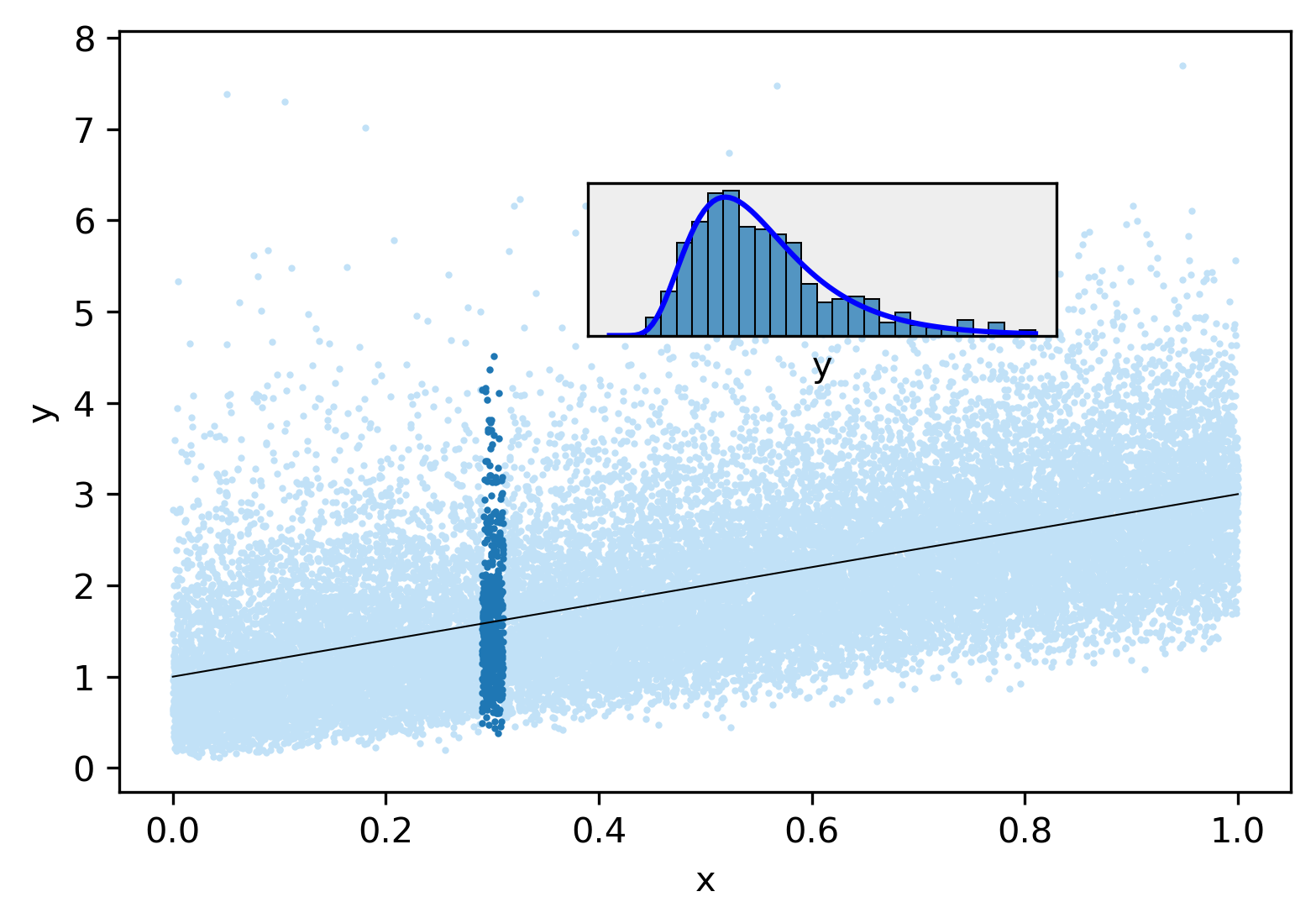}
  \caption{A synthetic data set containing $25,000$ sample $(x, y)$ pairs. Unlike the data in Figures~\ref{bg_01} and \ref{bg_02}, the $y$ values in this data set are strictly positive. The data in the range $0.29 < x < 0.31$ are highlighted. The inset shows the asymmetric histogram for the $y$ values in this subset along with the computed best-fit lognormal probability density function. The thin black line shows the conditional expected value of $y$ given $x$.}
  \label{bg_03}
\end{figure}

Figure~\ref{bg_03} shows a synthetic data set containing $25,000$ sample $(x, y)$ pairs. Unlike the data in Figures~\ref{bg_01} and \ref{bg_02}, the $y$ values in this data set are strictly positive. The histogram of the $y$ values for the subset defined by the range $0.29 < x < 0.31$ is not symmetric and would not be well modeled by a normal distribution. An asymmetric, strictly positive, distribution like the lognormal or gamma distribution could be a more appropriate choice.

The choice of the form for the conditional distribution is an important modeling decision that employs domain-specific knowledge and requires diagnostic checks after training. This topic will be explored more thoroughly in Section~\ref{S_form}.

\subsection{Heteroscedastic}
\begin{figure}[!htb]
  \centering
  \includegraphics[width=3.5in]{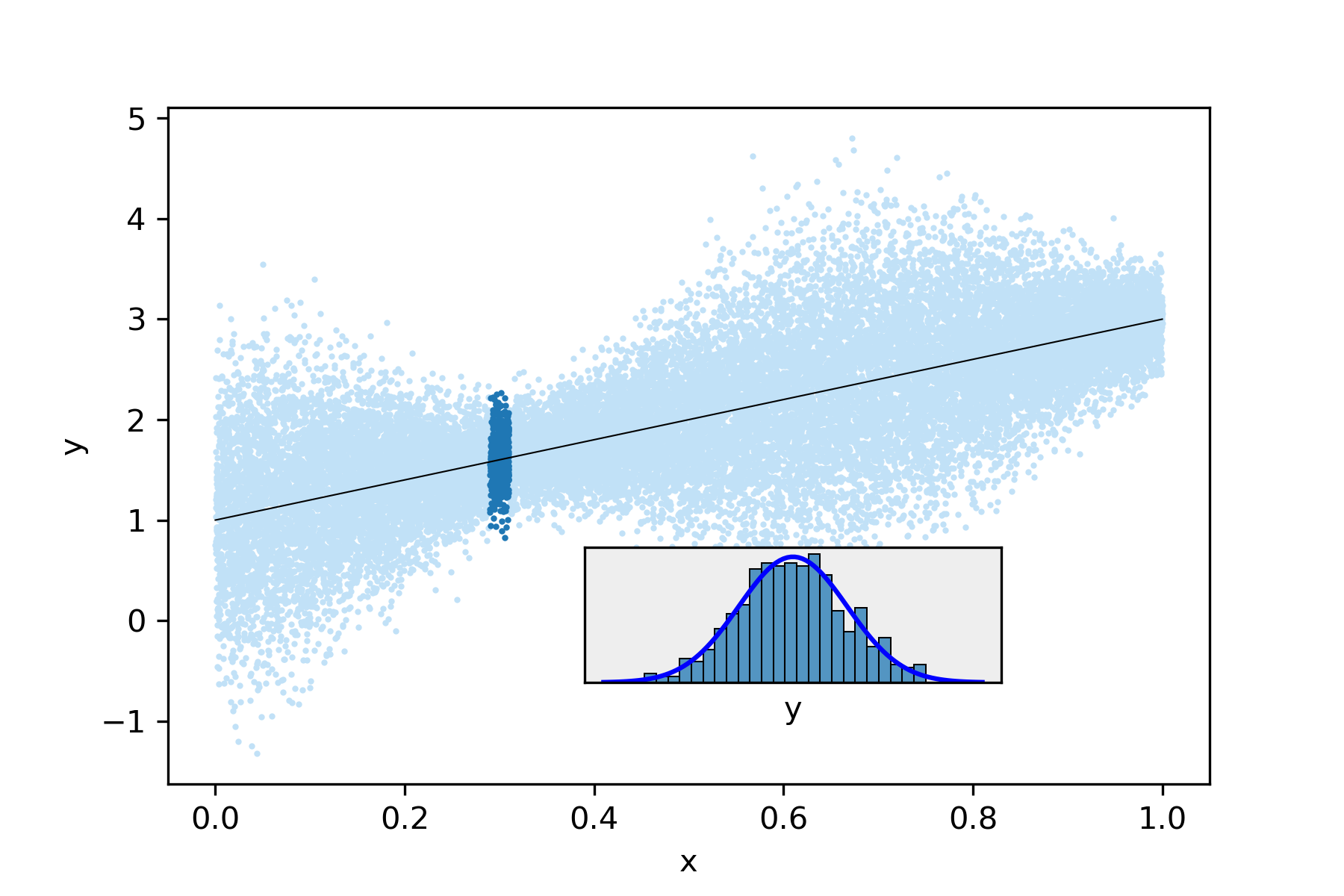}
  \caption{A synthetic data set containing $25,000$ sample $(x, y)$ pairs. These data are heteroscedastic. The data in the subset, range $0.29 < x < 0.31$, are highlighted. The inset shows the histogram for the $y$ values in this subset along with the computed best-fit normal probability density function. The thin black line shows the conditional expected value of $y$ given $x$.}
  \label{bg_04}
\end{figure}

Compare the data set shown in Figures~\ref{bg_02} and \ref{bg_04}. The mean trends of the two data sets are identical. The conditional distributions of $y$ around $x = 0.3$ are both reasonably well modeled by normal distributions. The two data sets are, however, fundamentally different.  

The data in Figure~\ref{bg_02} are {\it homoscedastic} (same variance, not a function of $x$). The data in Figure~\ref{bg_04} are {\it heteroscedastic} (changing variance as a function of $x$).  The spread of the $y$ values in Figure~\ref{bg_04} is a function of the $x$ values.  

In almost all applications of curve-fitting software the $y$ data are modeled as homoscedastic. When we use a simple least squares (or root-mean-square error) loss function, we implicitly model the $y$ data as homoscedastic. However, there are many circumstances where a heteroscedastic model could be a better choice: for example, the data in Figure~\ref{bg_04}.

\subsection{Nonlinear}
\begin{figure}[!htb]
  \centering
  \includegraphics[width=3.5in]{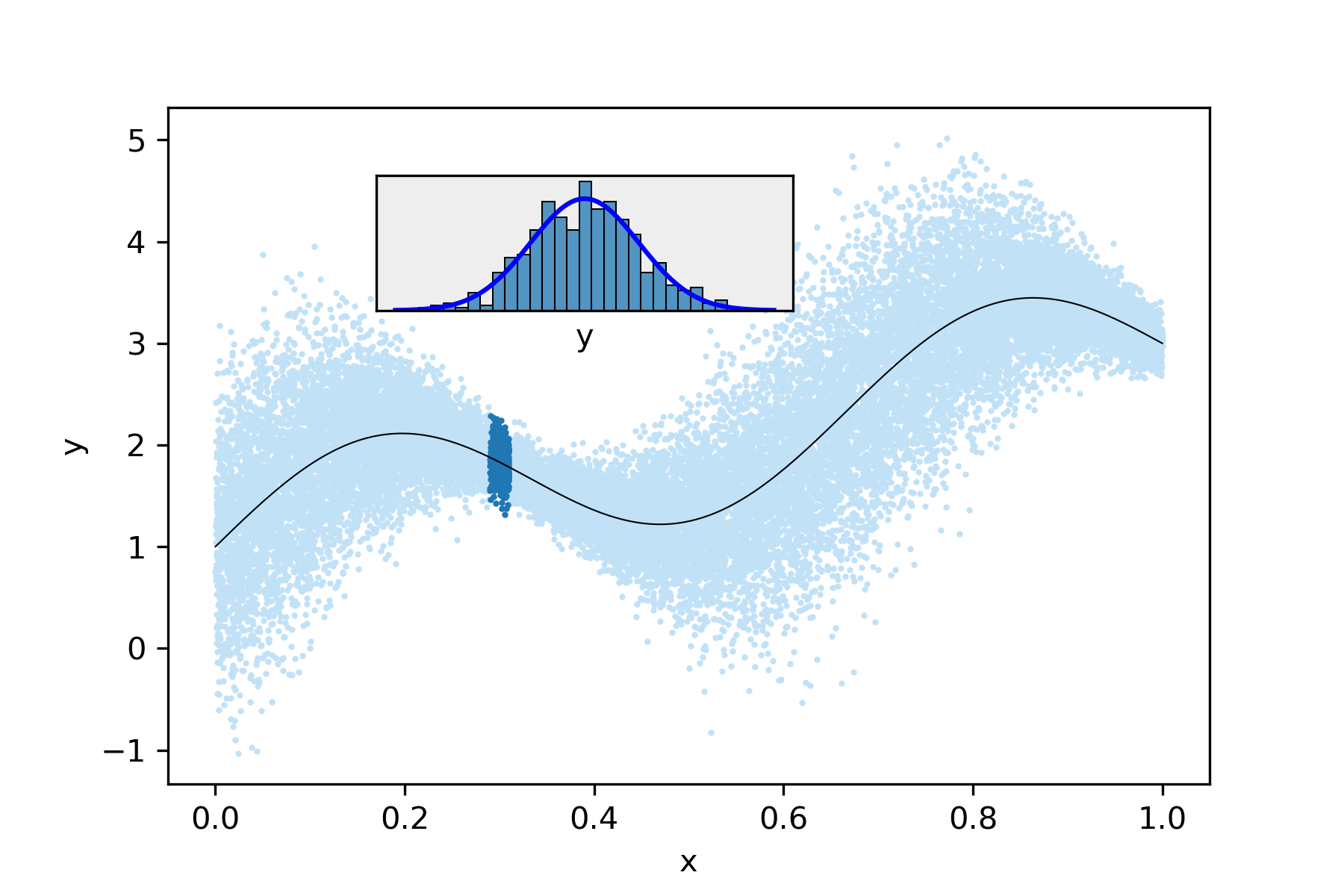}
  \caption{A synthetic data set containing $25,000$ sample $(x, y)$ pairs. These data are heteroscedastic and nonlinear. The data in the subset, range $0.29 < x < 0.31$, are highlighted. The inset shows the histogram for the $y$ values in this subset along with the computed best-fit normal probability density function. The thin black line shows the conditional expected value of $y$ given $x$.}
  \label{bg_05}
\end{figure}

The mean trends of the data sets shown in Figures~\ref{bg_01} through \ref{bg_04} are identical. The conditional expected value of $y$ is a simple linear function of $x$. By contrast, the conditional expected value of $y$ for the data shown in Figure~\ref{bg_05} is a {\it nonlinear} function of $x$. Nonetheless, the conditional distribution of the $y$ values in Figure~\ref{bg_05} is well modeled by a normal distribution, as shown in the inset for $0.29 < x < 0.31$.

\subsection{Multivariate}
 We are interested in use cases where $x$ is a vector and not a scalar: {\it multivariate} data.  The vector \textbf{x} could be bivariate (e.g. eastings and northings) or highly multivariate (e.g. the vectorization of the $180 \times 360$ one-degree cells in a global circulation model). The nonlinear heteroscedastic parametric regression outlined above applies equally well to multivariate data sets.

\section{Selecting the Form of the Conditional Distribution}\label{S_form}
The choice of the form for the conditional distribution is an important modeling decision that employs domain-specific knowledge.

\subsection{Should not be based on the histogram}
\begin{figure}[!htb]
  \centering
  \includegraphics[width=3.5in]{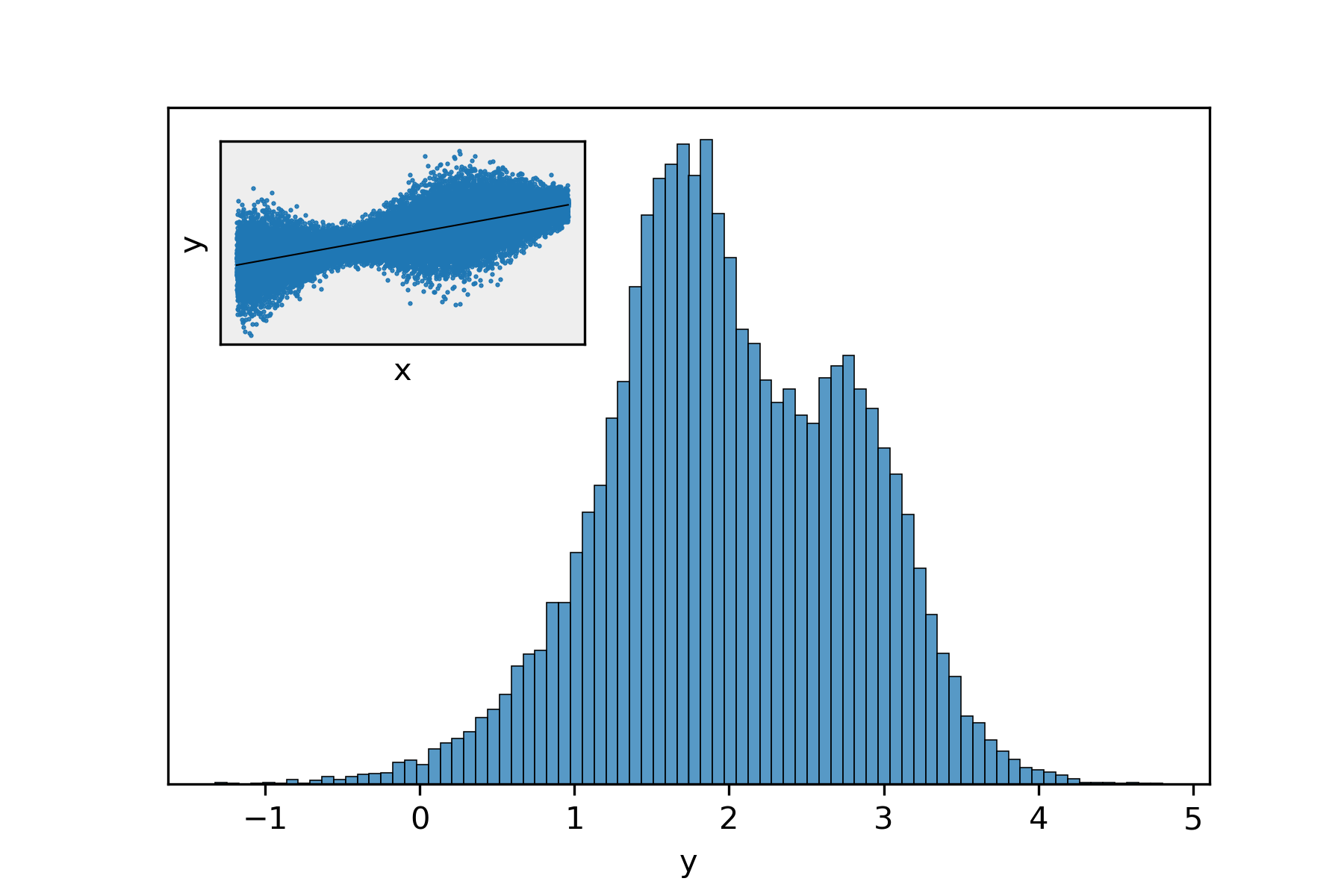}
  \caption{The histogram for all of the $y$ values in the data set shown in Figure~\ref{bg_04}. This histogram is not well modeled as a normal distribution. The inset shows the Figure~\ref{bg_04} data.}
  \label{bg_06}
\end{figure}

Reconsider the synthetic data set shown in Figure~\ref{bg_04}. The histogram for all of the $y$ values in this data set is shown in Figure~\ref{bg_06}. This histogram is bimodal and asymmetric, with a longer tail to the left than to the right. This histogram is not well modeled as a normal distribution. Nonetheless, the conditional distribution of $y$ given $x$ is a normal distribution.

This example demonstrates why we cannot simply select the form of the conditional distribution for $y$ by looking at the full distribution of $y$ (i.e. the apparent marginal distribution of $y$).  

\subsection{The valid range of $y$}
To choose the form of the conditional distribution, we ask: what is the valid range of $y$? Our answer may be partially informed by the data at hand\footnote{See Section~\ref{PIT} for one approach.}, but it must also incorporate our knowledge of the data: the data source, measuring devices, recording techniques, and representation.

If we anticipate that $y$ spans a range of positive and negative values, then a conditional probability distribution with support from $-\infty$ to $\infty$ may be appropriate.  The normal is an example of such a distribution (a flexible alternative is presented in Section~\ref{shash}.).

If our $y$ values are strictly positive, than a conditional probability distribution with support from $0$ to $\infty$ may be more appropriate. The lognormal and gamma are common examples of such distributions used in the geosciences (e.g. \cite{Foster2006,MartinezVillalobos2019}).

If our $y$ variable denotes a percentage or proportion, then a distribution with finite support from $0$ to $1$ may be more appropriate.  The beta and Kumaraswamy are two such distributions that have been used to model hydrological variables such as daily rainfall (e.g. \cite{Mielke1975,Kumaraswamy1980}).

If our $y$ values denote orientation and are constrained to the interval $0$ to $2\pi$, then a circular conditional probability distribution may be more appropriate. A circular distribution properly wraps around at $2\pi$. The von Mises is an example of such a distribution that has been used to model wind direction (e.g. \cite{Justus1978,Carta_2008}).  

\section{Neural Networks for Parameter Estimation }\label{S_loss}
In this note, we train a neural network to predict the \textit{local} uncertainty of $y$ by predicting the parameters of the conditional distribution of $y$ as a function of multivariate inputs $\textbf{x}$. As discussed in Section~\ref{S_form}, we must choose the form of the conditional distribution. 

\subsection{Predicting parameters of the sinh-arcsinh normal distribution}\label{shash}
\begin{figure}[!htb]
  \centering
  \includegraphics[width=3.5in]{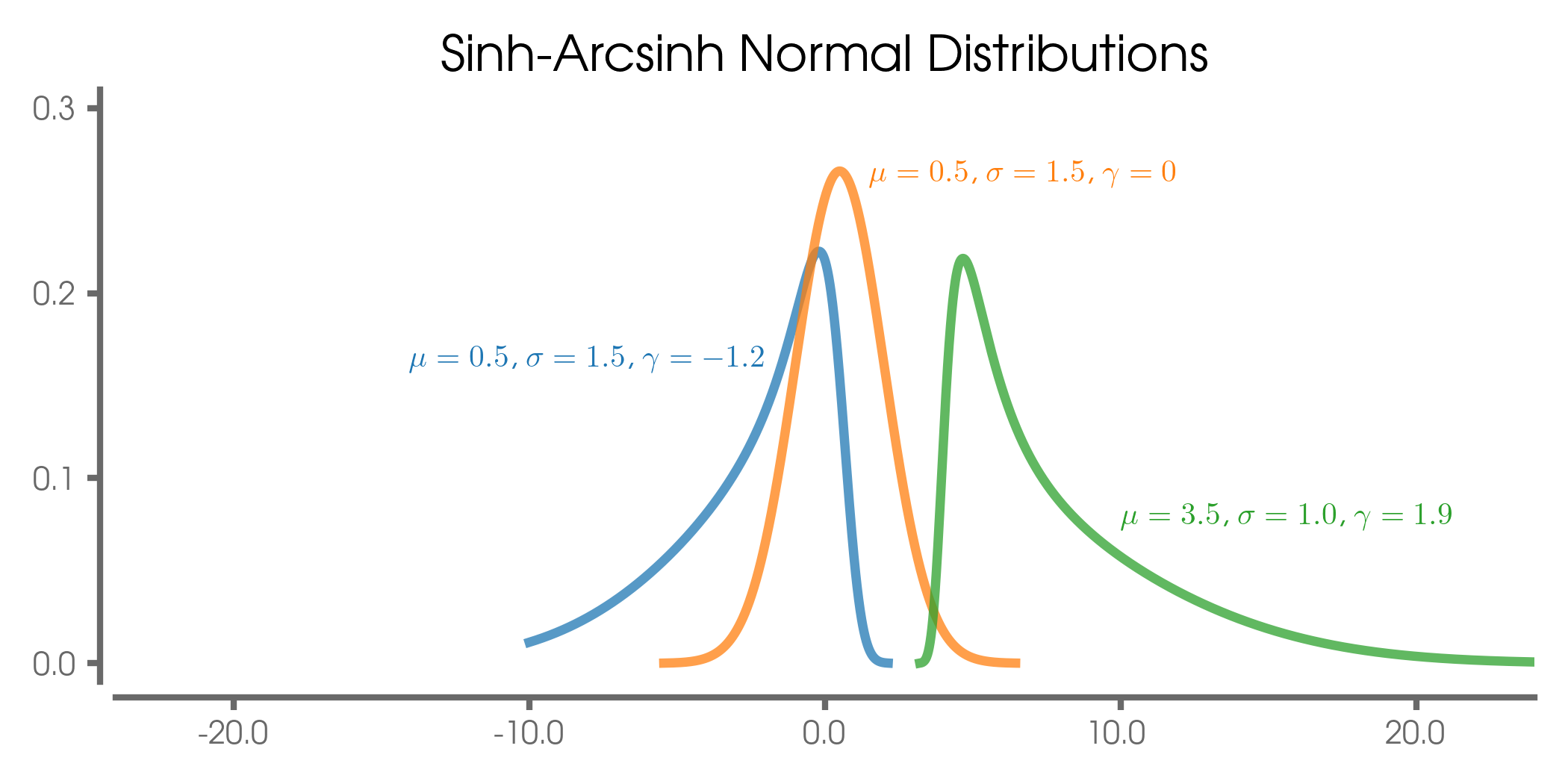}
  \caption{Example probability distributions for the family of sinh-arcsinh normal distributions for different parameter values. In all cases, the tailweight $\tau = 1$.}
  \label{fig_sas}
\end{figure}

Here, we demonstrate the use of the \textbf{sinh-arcsinh normal distribution} \cite{Jones2009,Jones2019,Duerr2020}, which takes four parameters in {\tt Tensorflow Probability}\footnote{The {\tt Tensorflow Probability} documentation for the sinh-arcsinh normal distribution lists functions for mean, standard deviation, and variance, but they are not yet implemented in the library (v. 0.13.0). Appendix~\ref{D_helper} gives code to fill this gap.}: location ($\mu$), scale ($\sigma$), skewness ($\gamma$), and tailweight ($\tau$). Examples of this distribution are shown in Figure~\ref{fig_sas}, where we set the tailweight $\tau=1$ for simplicity and to reduce the number of predicted parameters. 

The sinh-arcsinh normal distribution provides a basis for representing heteroscedastic and asymmetric uncertainties (Section~\ref{S_back}) in an intuitive manner \cite{Duerr2020}, of which the normal distribution is a special case (skewness $\gamma=0$, tailweight $\tau=1$; orange curve in Figure~\ref{fig_sas}).

\begin{figure}[!htb]
  \centering
  \includegraphics[width=2.0in]{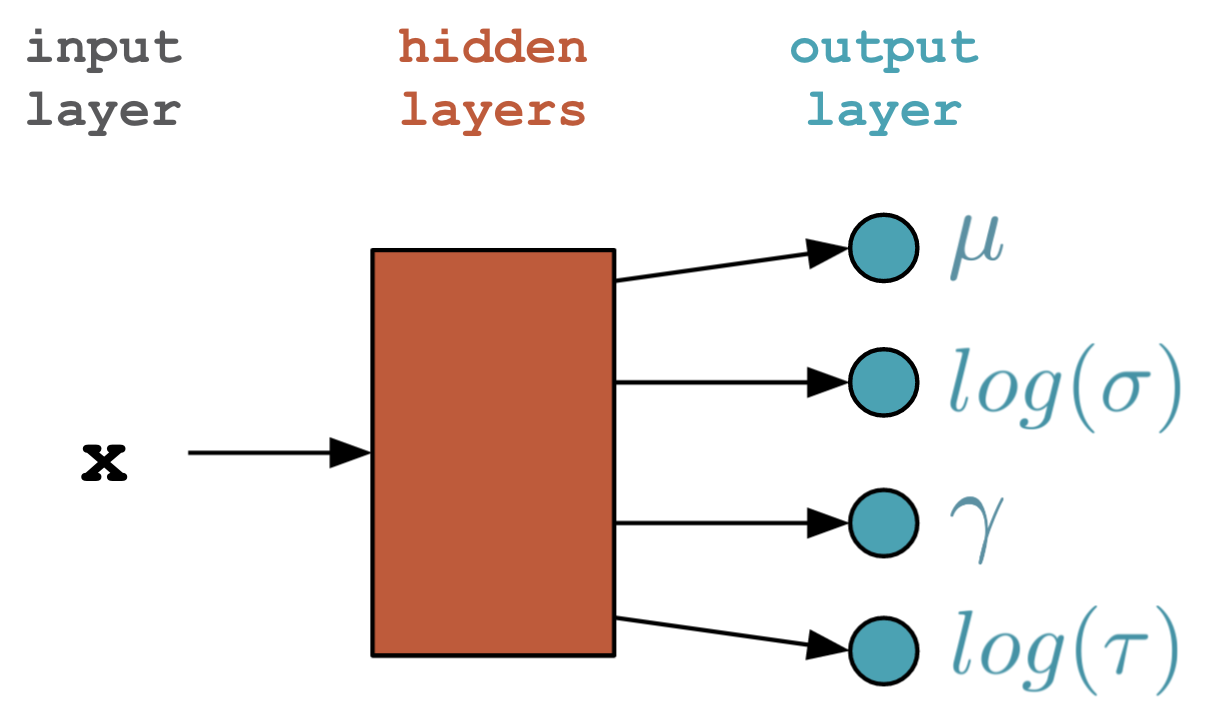}
  \caption{A generic schematic showing a neural network setup with an input layer and hidden layers that ultimately predict four parameters of the sinh-arcsinh normal distribution.}
  \label{fig_arch}
\end{figure}

The neural network is tasked with predicting the four parameters of the sinh-arcsinh normal distribution (Figure~\ref{fig_arch}). That is, the output of the network contains four units capturing $\mu, \sigma$, $\gamma$ and $\tau$. 

The parameters $\sigma$ (scale) and $\tau$ (tailweight) must be strictly positive. To accomplish this, we follow a common trick in computer science \cite{Duerr2020} to ensure that $\sigma$ and $\tau$ are always strictly positive by tasking the network to predict the $\log{\sigma}$ and the $\log{\tau}$ rather than $\sigma$ and $\tau$ themselves (Figure~\ref{fig_arch}). This trick is discussed further in Appendix~\ref{A_constraints}.

In the examples shown here, we predict the four parameters of the sinh-arcsinh normal distribution using fully connected neural networks. However, a benefit of this general approach for including uncertainty is that it can be easily applied to most any neural network architecture, as it only requires modification of the output layer and the loss function (see next section).

\subsection{The loss function}
The neural network architecture shown in Figure~\ref{fig_arch} trains using the negative log-likelihood loss defined for the $i^{th}$ input sample, $x_i$, as
\begin{equation}
\mathcal{L}(x_i) = -\log p_i. \label{loss_base}
\end{equation}
where $p_i$ is the value of the sinh-arcsinh normal probability density function ($\mathcal{P}$) with predicted parameters $(\mu,\sigma,\gamma,\tau)$ evaluated at the true label $y_i$. That is, 
\begin{equation}
p_i = \mathcal{P}(y_i | \mu,\sigma,\gamma,\tau) 
\end{equation}
Example {\tt Tensorflow} code is provided in Appendix~\ref{D_loss}.

\begin{figure}[!htb]
  \centering
  \includegraphics[width=5.5in]{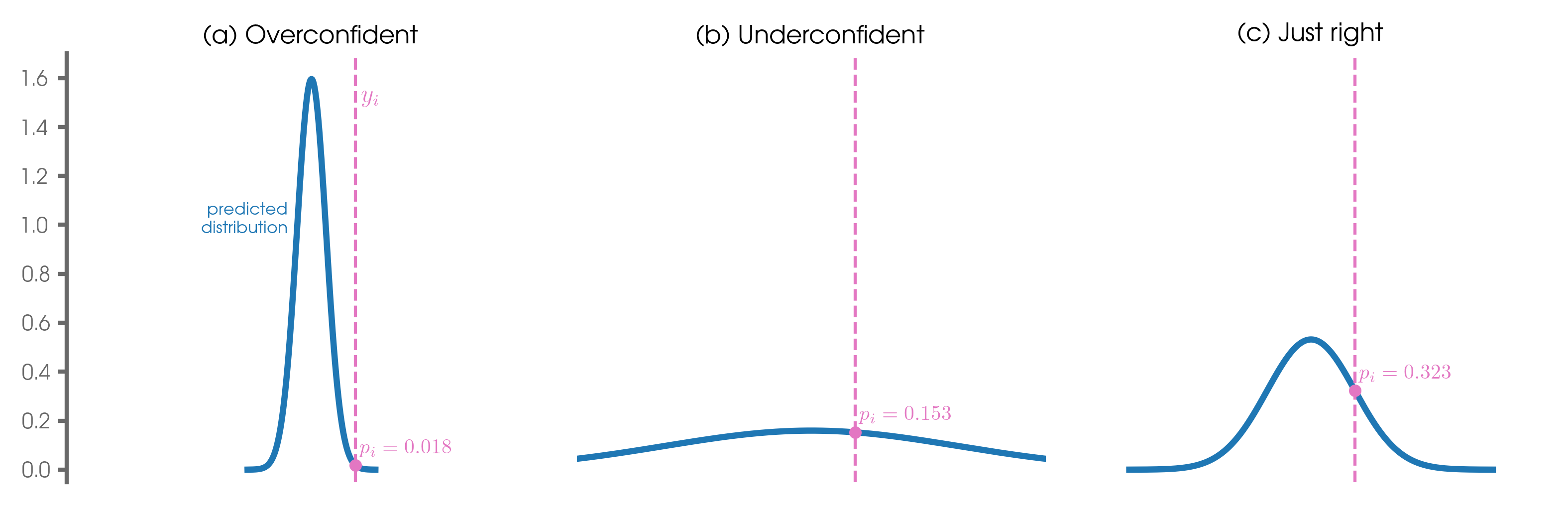}
  \caption{Schematic showing how the loss function probability $p_i$ is dependent on the predicted probability distribution.}
  \label{fig_goldy}
\end{figure}

Figure~\ref{fig_goldy} shows schematically how the probability $p_i$ depends on the specific parameters of the distribution predicted by the network. For example, if the predicted distribution is overconfident (the distribution's width is too small), $p_i$ will be small and the loss $\mathcal{L}$ will be large (Figure~\ref{fig_goldy}a). If the network is underconfident (the distribution's width is too large), $p_i$ will also be small and the loss $\mathcal{L}$ will once again be large (Figure~\ref{fig_goldy}b). Like Goldilocks, the network needs to get things just right, by properly estimating its uncertainty to obtain its best guess. Better predictions are associated with larger $p_i$, and thus smaller loss $\mathcal{L}$ (Figure~\ref{fig_goldy}c). Note that this approach is directly tied to the standard method of parameter estimation in statistics known as \textit{Maximum Likelihood Estimation} \cite{Duerr2020}.

It is important to stress that while the following examples use the sinh-arcsinh normal probability density function, the loss function in Eq.~\ref{loss_base} can be used for any distribution. The trick to training the neural network is that the loss for observation $i$ is just the predicted probability density function evaluated at the observe $y_i$.

\section{Demonstration with Simple 1D Data}\label{1D_data}
\subsection{Heteroscedastic, symmetric uncertainties}\label{1D_symm}
\begin{figure}[!htb]
  \centering
  \includegraphics[width=5.5in]{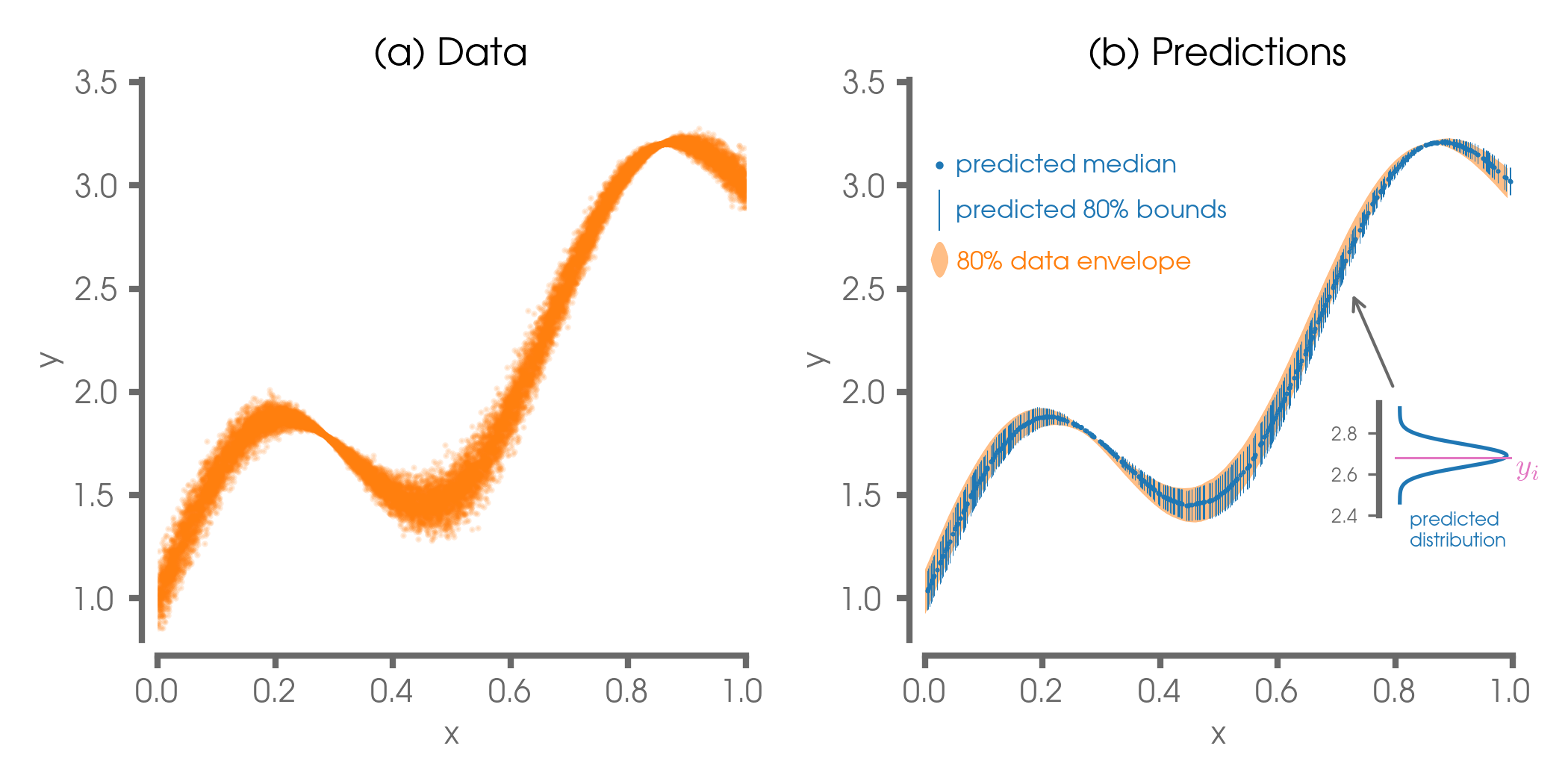}
  \caption{Predictions by the neural network with heteroscedastic, normal uncertainties.}
  \label{fig_norm}
\end{figure}

We first demonstrate the utility of this method by introducing the data set shown in Figure~\ref{fig_norm}a. The uncertainty in $y_i$ for each value of $x_i$ is defined by symmetric, normally distributed noise that is a function of $x$ itself. That is, the noise is heteroscedastic, but locally symmetric. We train a fully connected neural network to predict $\mu, \sigma, \gamma$ of the sinh-arcsinh normal distribution, setting $\tau=1$ for simplicity. The network has three hidden layers of 50 units each using the {\tt ReLU} activation, a learning rate of 0.0001, and the {\tt Keras SGD} optimizer. We train the model on 20,000 training samples, using early stopping based on the loss of 2,500 validation samples to optimize the number of training epochs. We then analyze an additional 2,500 samples as testing data. Example {\tt Tensorflow} code is provided in Appendix~\ref{C_network}.

Once training is complete, we evaluate the success of the neural network based on 2,500 samples included in a separate testing set. For each sample in the testing set, the neural network predicts the three free parameters of the sinh-arcsinh normal distribution.  We plot the 80\% prediction interval bounds of each of those predicted distributions as a function of $x$ in Figure~\ref{fig_norm}b. The neural network's predicted uncertainty bounds appear to align  well with the true bounds of the data (orange shading), becoming larger for certain values of $x$ and smaller for others. Furthermore, the network learns that the residuals are locally symmetric and thus predicts near-zero skewness (i.e. $\gamma < 0.03$; not shown). 

\subsection{Heteroscedastic, asymmetric uncertainties}\label{1D_asymm}
\begin{figure}[!htb]
  \centering
  \includegraphics[width=5.5in]{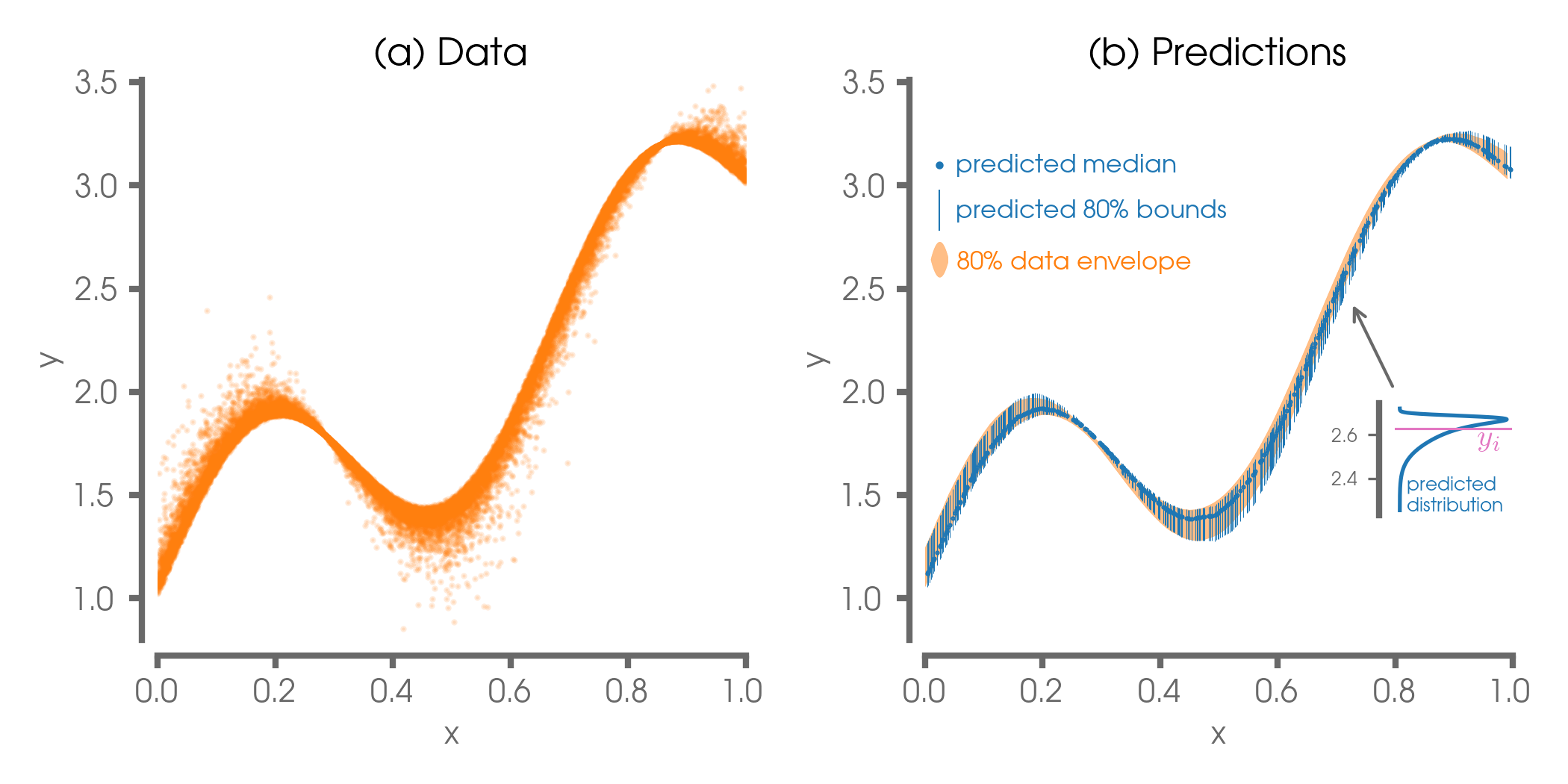}
  \caption{Predictions by the neural network with heteroscedastic, asymmetric uncertainties.}
  \label{fig_lognorm}
\end{figure}

The sinh-arcsinh normal distribution permits the consideration of uncertainties with both heteroscedastic and asymmetric uncertainties, e.g. rainfall rates that are constrained to be non-negative. Figure~\ref{fig_lognorm} shows a similar example to Figure~\ref{fig_norm}, but this time the uncertainties are skewed in either the positive or negative direction as a function of $x$. In this case, the neural network predicts non-zero skewness values and, once again, is able to capture the true 80\% confidence interval of the data (Figure~\ref{fig_lognorm}b).

\section{Some Model Diagnostics}\label{S_diag}
After we have trained our network, what evidence do we have that our model is any good? On the one hand, one could just compare the true $y$ value with the mean or median of the predicted conditional distribution. But which do you choose --- the median or the mean? Furthermore, this type of residual diagnostic does not address whether the predicted uncertainties (i.e. distributions) are useful. Any neural network can spew out estimations of parameters --- it is up to the user to determine whether they hold any meaning. So how do we do that?

\subsection{Visual inspection}
In the case of 1D data, like the data analyzed in Section~\ref{1D_data}, we graph the testing data on an $(x, y)$ plot and overlay the associated prediction intervals (see Figures~\ref{fig_norm} and \ref{fig_lognorm}). If the  $80\%$ prediction interval envelope captures about $80\%$ of the data, then our confidence in the model increases. 

Unfortunately, this visual approach does not translate to multivariate cases. We cannot plot hyper-dimensional data on a computer screen.

\subsection{Standardized residuals}
\begin{figure}[!htb]
  \centering
  \includegraphics[width=3.5in]{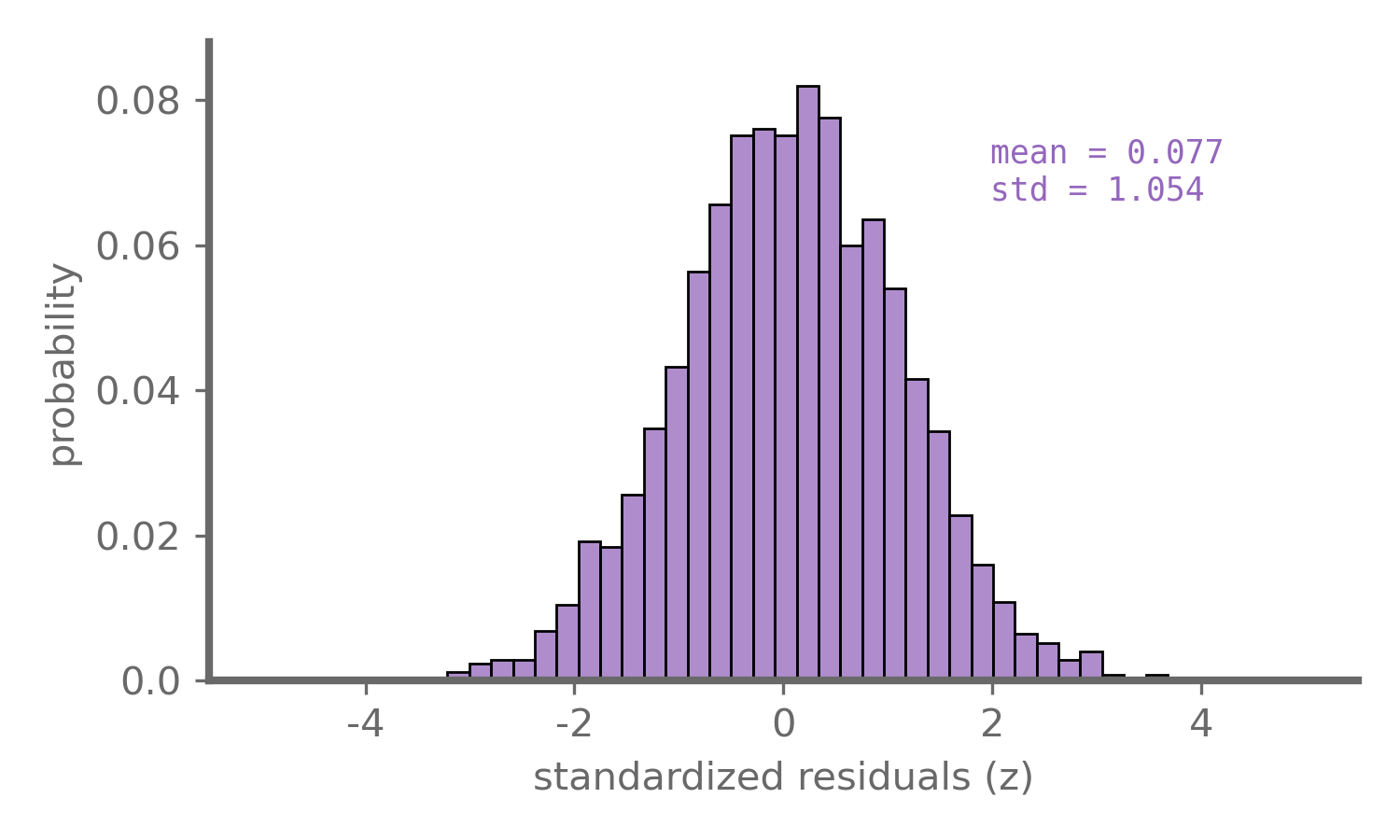}
  \caption{Histogram of standardized residuals of the testing set from the case of heteroscedastic, symmetric uncertainties examined in Section~\ref{1D_symm}. The mean is about 0, the standard deviation is about 1, and the distribution is normal (a bell curve).}
  \label{fig_hist_sym}
\end{figure}

\begin{figure}[!htb]
  \centering
  \includegraphics[width=3.5in]{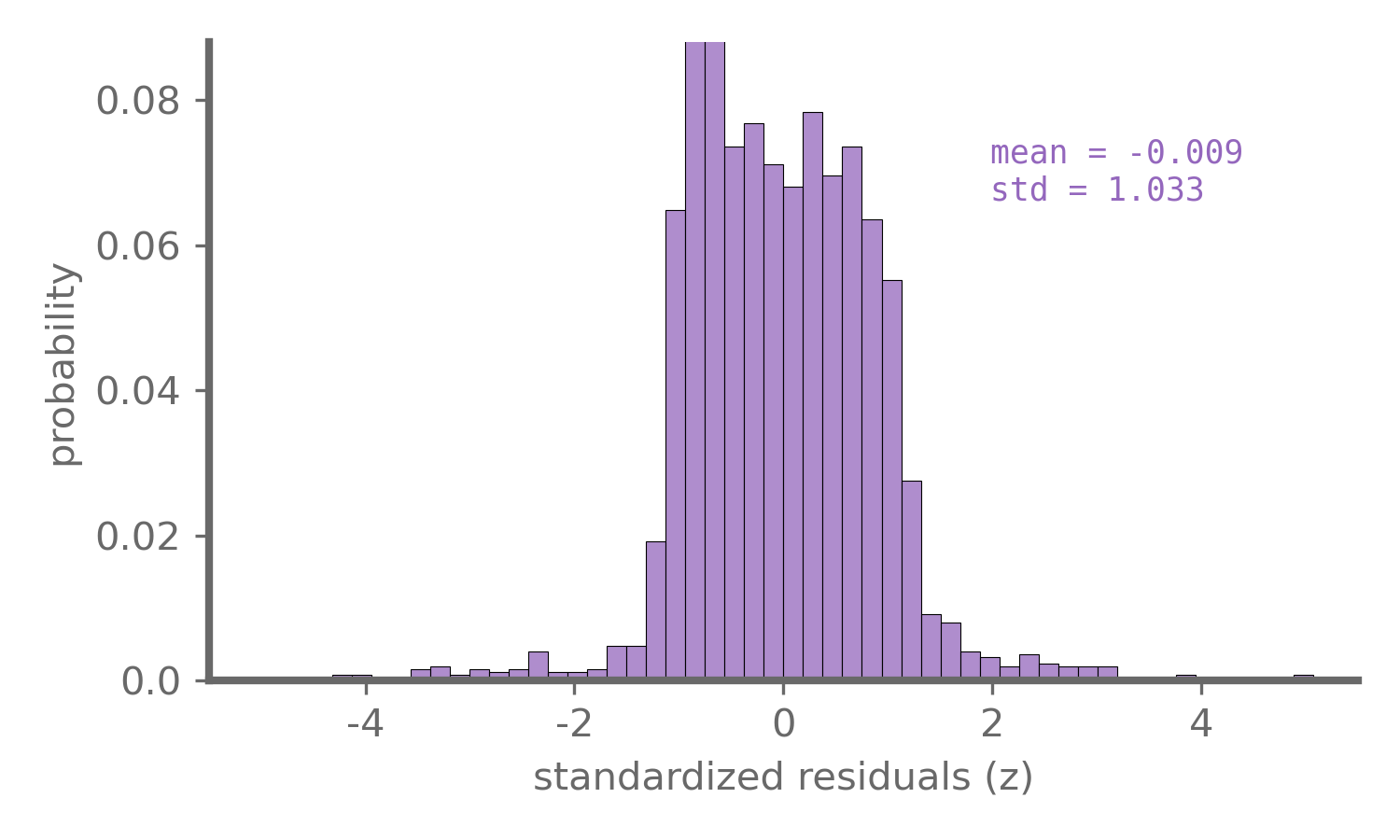}
  \caption{Histogram of standardized residuals of the testing set from the case of heteroscedastic, asymmetric uncertainties examined in Section~\ref{1D_asymm}. The mean is about 0, the standard deviation is about 1, and the distribution is decidedly not normal.}
  \label{fig_hist_asym}
\end{figure}

Figures~\ref{fig_hist_sym} and \ref{fig_hist_asym} show another way to visualize the predicted uncertainty; this works for multivariate data. We compute the standardized residual for each prediction of the testing set, defined as:
\begin{equation}
z_i = \frac{y_i - \bf{mean}(y|x_i)}{\bf{std}(y|x_i)}
\end{equation}
where ${\bf{mean}}(y|x_i)$ and ${\bf{std}}(y|x_i)$ are the mean and standard deviation of the conditional distribution at $x_i$ predicted\footnote{If the normal distribution is chosen as the modeled conditional distribution, then ${\bf{mean}}(y|x_i)$ and ${\bf{std}}(y|x_i)$ are the two predicted parameters from the network for sample $i$.  For some conditional distributions, the predicted parameters are not the mean and standard deviation (e.g. the lognormal distribution is usually parameterized by the log-mean and log-standard deviation).  In such cases, the mean and standard deviation must be computed from the predicted parameters before one computes the standardized residuals.} by the network. If the mean and standard deviation of the modeled conditional distributions are approximately correct, the mean and standard deviations of the $z$'s should be approximately $0$ and $1$.  See, for example, \cite[Section 19.3]{Biecek2021}.

The distribution of standardized residuals in Figures~\ref{fig_hist_sym} and \ref{fig_hist_asym} have means of about $0$ and standard deviations of about $1$. These observations further support our assertion that the network's predictions can be interpreted as approximate conditional probability distributions.

We observe the difference in shapes for the two histograms.  Figure~\ref{fig_hist_sym} appears to be normal (a bell curve), and Figure~\ref{fig_hist_asym} is decidedly not normal.  This difference is a consequence of how we created the data. We cannot conclude that the first model is better than the second, because its histogram of standardized residuals is ``more normal."

\subsection{Sign test}
The sign test checks how well the model predicts the middle of the conditional distribution. Let $\bf{med}(y|x_i)$ denote the median of the of the predicted\footnote{For most conditional distributions, the median will need to be computed from the predicted parameters.} conditional distribution for sample $i$.  Let
\begin{itemize}
    \item $n_p$ denote the number of times that $y_i > \bf{med}(y|x_i)$, and
    \item $n_n$ denote the number of times that $y_i < \bf{med}(y|x_i)$.
\end{itemize}

The trained model for the heteroscedastic, symmetric data set discussed in Section~\ref{1D_symm} yields $n_p = 1215$ and $n_n = 1285$ for the $2,500$ test samples. Since the probability of randomly falling above, or below, the median is $50\%$, our results are equivalent to flipping a fair coin $2,500$ times and counting $1215$ heads and $1285$ tails.  Using a binomial distribution calculator\footnote{See, for example, \url{https://stattrek.com/online-calculator/binomial.aspx}.} (with $p = 0.5$, $n = 2500$, and $x = 1215$), we compute the probability of getting $1215$ or fewer heads out of $2500$ flips to be about $8\%$. This is not a low-probability event.  

Comparable calculations for the heteroscedastic, asymmetric data set discussed in Section~\ref{1D_asymm} yields $n_p = 1252$ and $n_n = 1248$ for the $2500$ test samples. Using a binomial distribution calculator (with $p = 0.5$, $n = 2500$, and $x= 1252$), we compute the probability of getting $1252$ or more heads out of $2500$ flips to be about $48\%$. This, too, is not a low-probability event.

These observations further support our assertion that the network's predictions can be interpreted as approximate conditional probability distributions. This diagnostic is called a {\it sign test} \cite[page 64]{Manoukian86}.  The sign test formally requires an assumption of independence, which is typically untestable and often false. Nonetheless, this diagnostic is exceedingly robust and surprisingly useful. See, for example, \cite[Section 3.4]{Conover1998}.

\subsection{{PIT} histogram}\label{PIT}
Dawid \cite{Dawid1984} introduced the probability integral transform (PIT) in 1984 as a calibration check for probabilistic forecasts. Consider sample $i$.  Our trained neural network gives us a conditional probability distribution for $y$ as a function of $x_i$. Denote the cumulative distribution function (CDF) for this conditional distribution $F(y|x_i)$. The probability integral transform (PIT) value for sample $i$, which we denote $p_i$, is 
\begin{equation}
    p_i = F(y_i | x_i)
\end{equation}
``In a nutshell, the PIT is the value that the predictive CDF attains at the observation'' \cite{Gneiting2014}. The histogram of all such value from a data set is called the PIT histogram. If the predictive model is ideal, then the PIT histogram follows a uniform distribution \cite{Gneiting2007}.

\begin{figure}[!htb]
  \centering
  \includegraphics[width=4.0in]{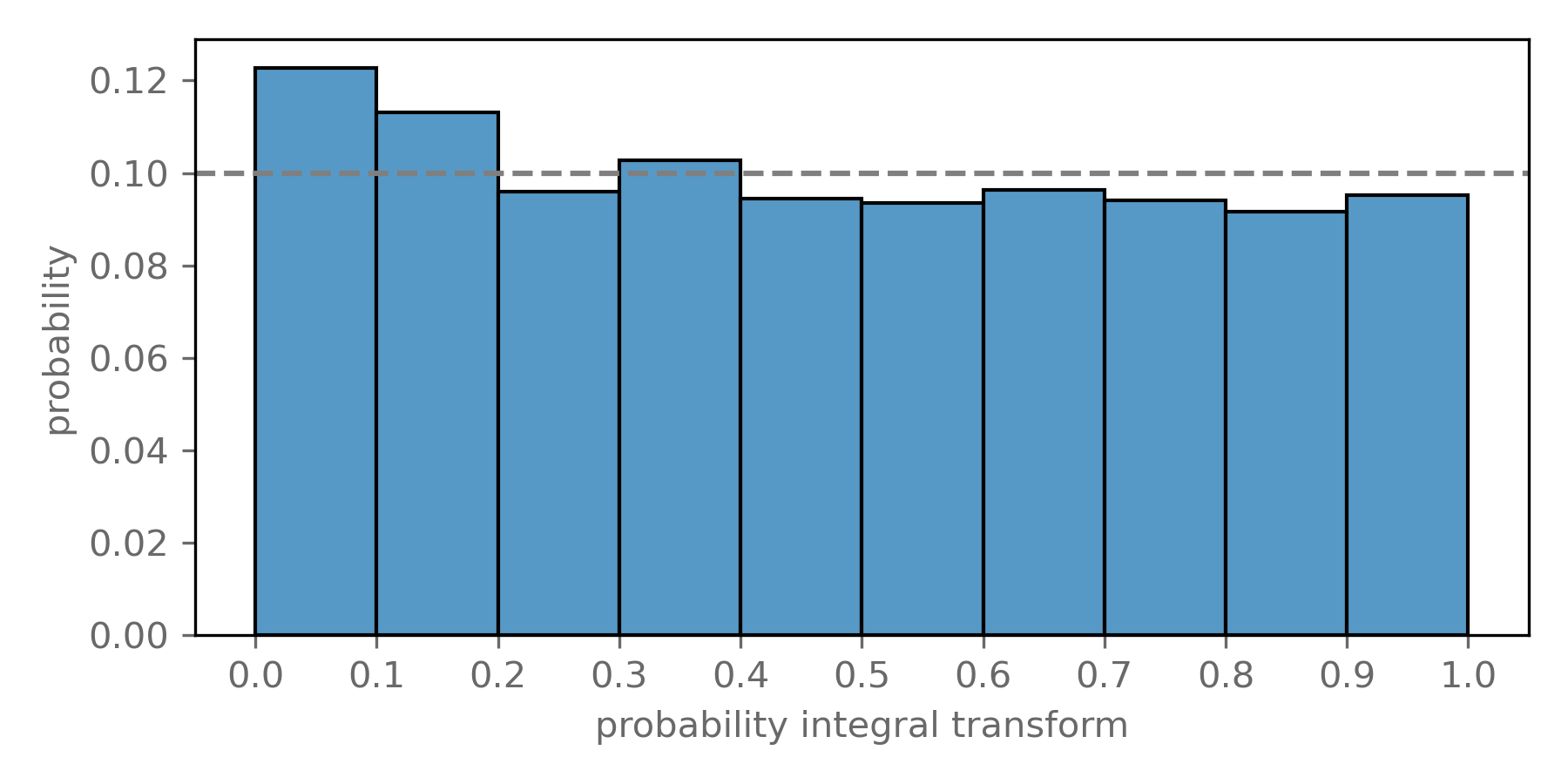}
  \caption{Probability integral transform (PIT) histogram of the testing set from the case of heteroscedastic, symmetric uncertainties examined in Section~\ref{1D_symm}. The horizontal dashed line denotes an equal probability of 0.1 across all bins.}
  \label{fig_pit_sym}
\end{figure}

Figure~\ref{fig_pit_sym} shows the PIT histogram for the test data using a trained model from the heteroscedastic, symmetric data set discussed in Section~\ref{1D_symm}. This histogram is almost uniform, but for the slight over representation in the first two bars. Our model is not capturing the behavior in the lower tail perfectly. 

\begin{figure}[!htb]
  \centering
  \includegraphics[width=4.0in]{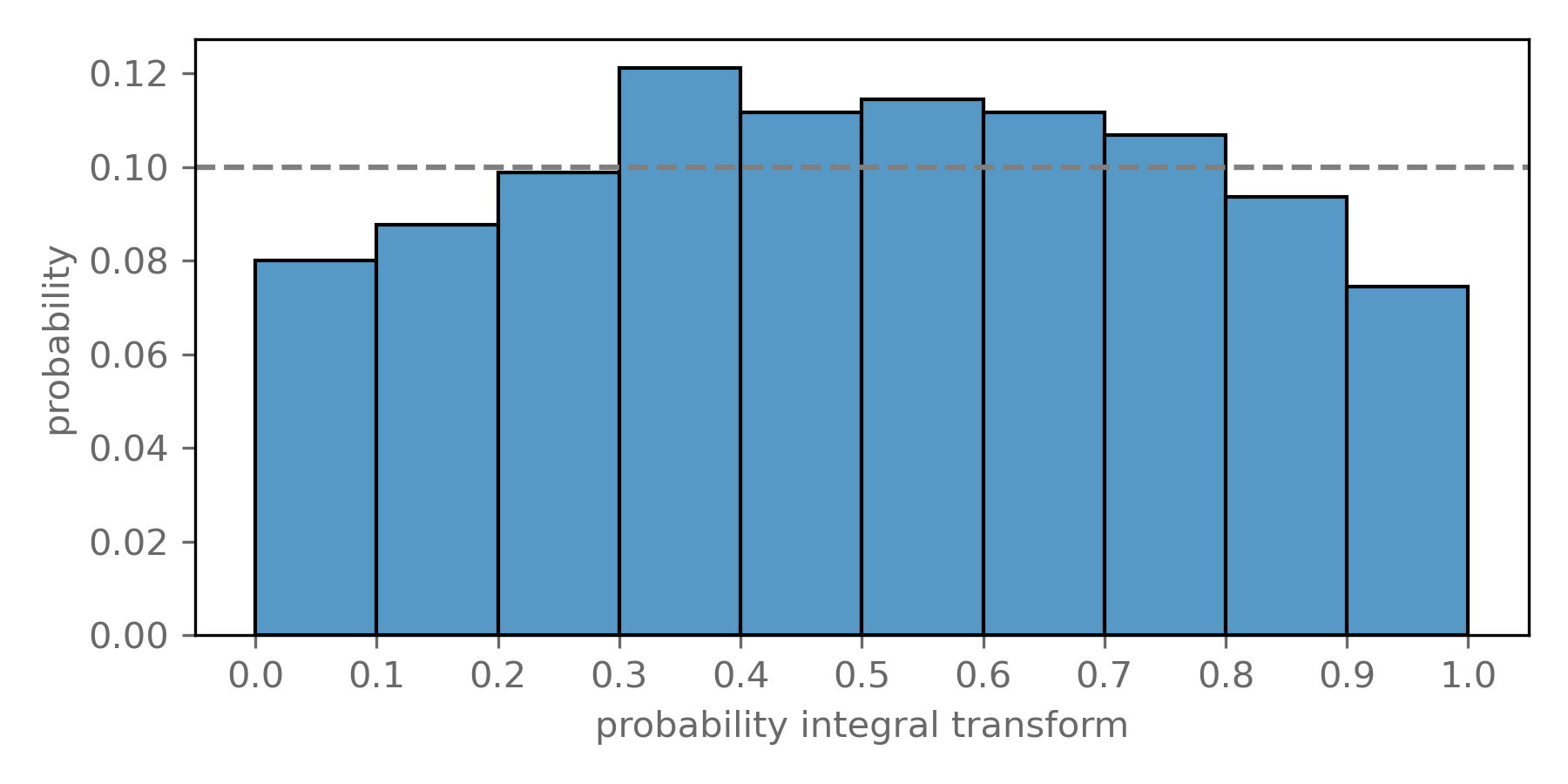}
  \caption{Probability integral transform (PIT) histogram of the testing set from the case of heteroscedastic, asymmetric uncertainties examined in Section~\ref{1D_asymm}. The horizontal dashed line denotes an equal probability of 0.1 across all bins.}
  \label{fig_pit_asym}
\end{figure}

Figure~\ref{fig_pit_asym} shows the PIT histogram for the test data using a trained model for the heteroscedastic, asymmetric data set discussed in Section~\ref{1D_asymm}. This histogram shows a distinct mound; the middle bars are taller than the ends. A PIT histogram with this pathology, mild though it is, suggests that the network may tend to be under-confident in its predictions (i.e. on average, the conditional distributions are too wide). 

As discussed in Section~\ref{S_form}, domain-specific knowledge and the data should both influence the selection of  the form of the conditional distribution.  The PIT histogram is a tool that we can use to combine both. Scheuerer and Möller \cite{Scheuerer_2015} use the PIT histogram to compare different predictive distribution models in an effort to identify the distribution type that yields the best calibration.

\clearpage
\section{Demonstration with Synthetic Climate Data}
Our examples thus far involve relatively simple 1D data sets. Therefore you may not be convinced it works in more realistic situations. Next, we move a step up in complexity and analyze the synthetic climate data set originally created by \cite{Mamalakis2021} and later used by \cite{BarnesBarnes2021}. In a nutshell, the data set consists of latitude-longitude maps (\textbf{x}) of synthetic sea-surface temperature (SST) anomalies and their associated labels $\hat{y}$. The $i$th label $\hat{y}_i$ is a continuous variable that is computed as the nonlinear sum of all of the SST anomalies in the $i$th map \textbf{x}$_i$. The exact form of the nonlinear function is explained in detail in \cite{Mamalakis2021}, but the most important thing is that it is known and exact.

\begin{figure}
  \centering
  \includegraphics[width=4.0in]{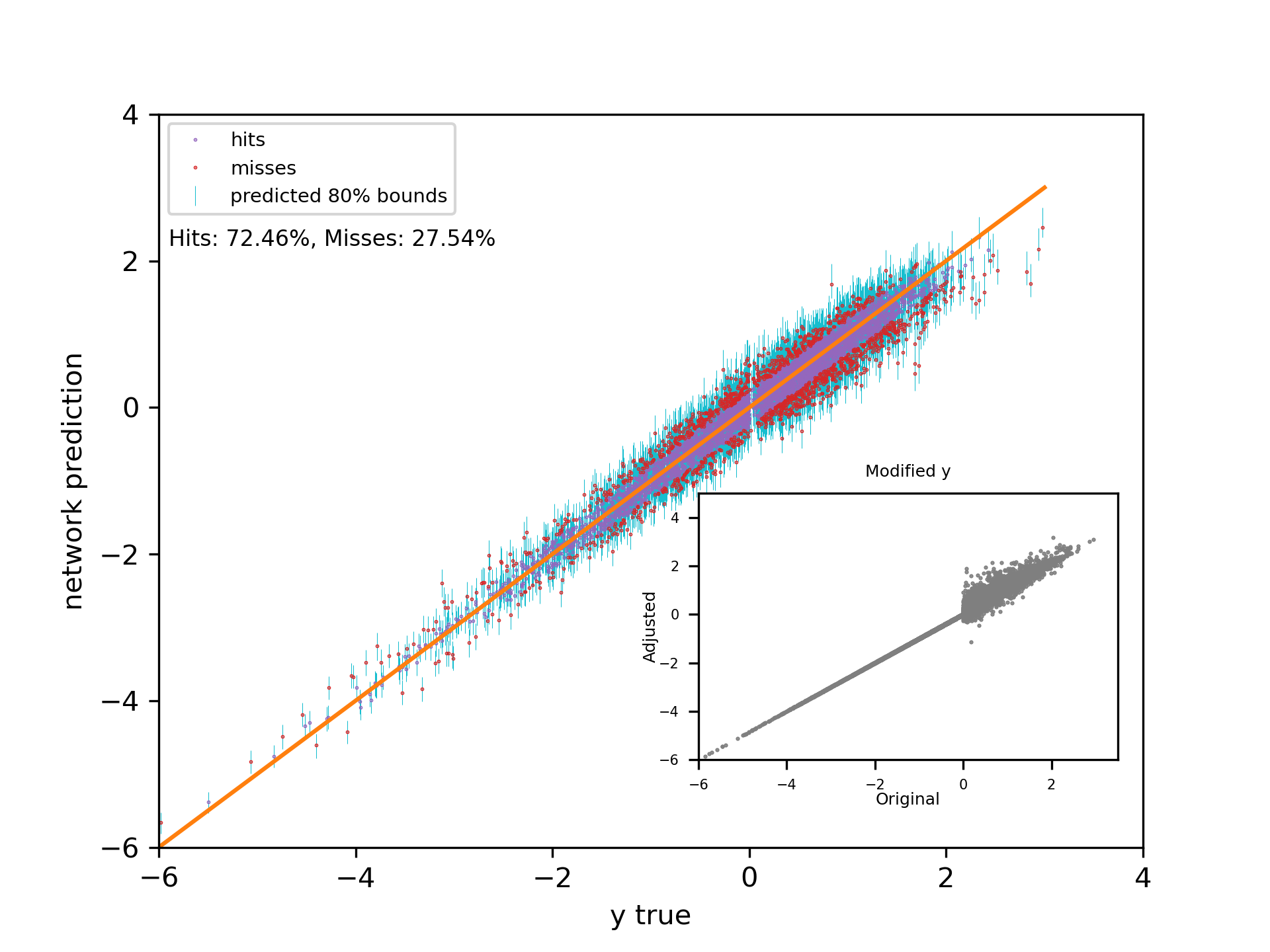}
  \caption{Performance of the neural network for the synthetic climate data. The blue vertical curve denotes the 80th percentile of the predicted distribution (i.e. 10th to 90th percentile of the distribution). Purple and red dots depict the medians of the distributions, where purple denotes that the true $y$ falls within the 80th percentile range (``hit'') and red denotes that it does not (``miss''). The inset shows the original $y$ labels and those adjusted by adding lognormal noise.}
  \label{fig_clim_log} 
\end{figure}

The synthetic data has a perfect, known nonlinear relationship between its input (maps of SST anomalies) and output $\hat{y}$. That is, there is no inherent uncertainty connecting the inputs to the outputs. We add uncertainty to the output by modifying the label $\hat{y}$ for particular samples. Specifically, for any sample with $\hat{y}>0$, we add lognormal noise such that 96\% of the adjusted $y$ values\footnote{We remove the hat symbol to denote the adjusted $y$ labels.} are skewed in the positive direction and the other 4\% are skewed in the negative direction. All $y$ labels less than $0$ remain untouched. The inset in Figure~\ref{fig_clim_log} depicts the original $\hat{y}$ labels on the x-axis and the adjusted $y$ labels on the y-axis. 

\begin{figure}
  \centering
  \includegraphics[width=2.5in]{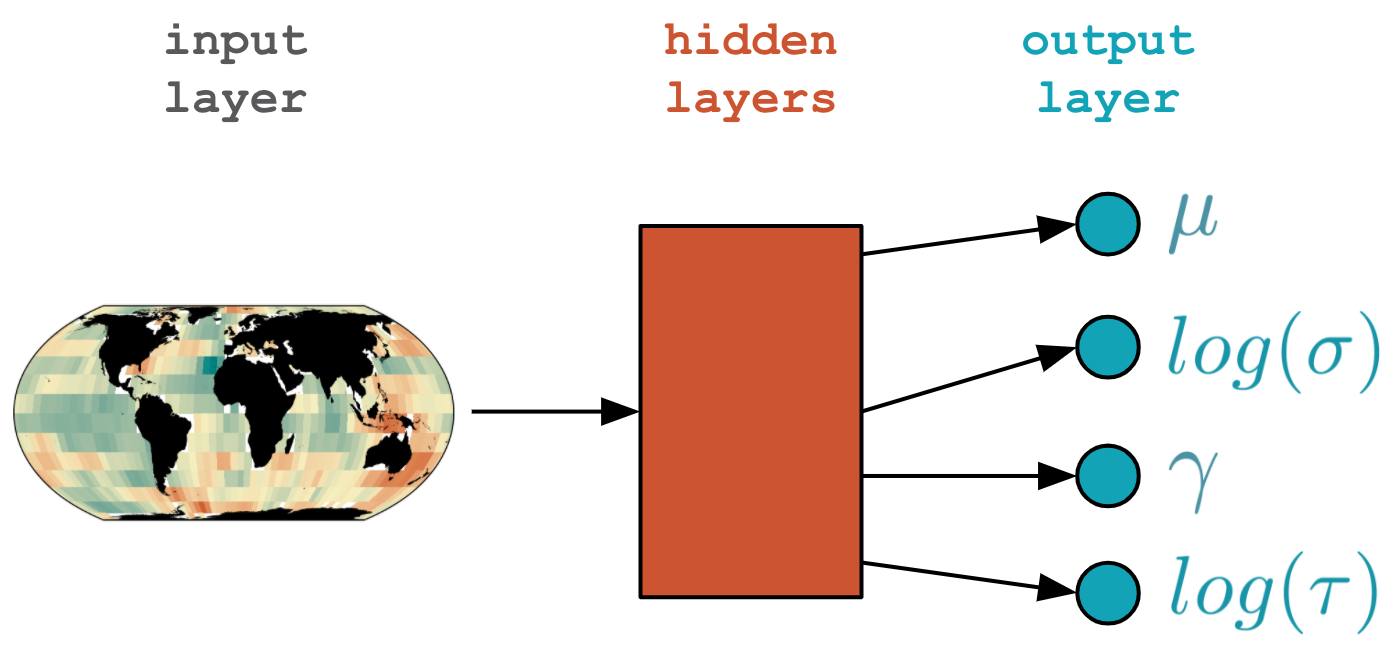}
  \caption{Schematic showing the neural network setup for the synthetic climate data. The network ingests a map of synthetic sea-surface temperature anomalies and is tasked with predicting the conditional distribution of the label $y$ in terms of the four parameters of the sinh-arcsinh normal distribution. }
  \label{fig_map} 
\end{figure}

We task a neural network with ingesting the maps of SST anomalies and predicting the conditional distribution of adjusted $y$ labels, as depicted in Figure~\ref{fig_map}. Specifically, we fit all four parameters ($\mu,\sigma,\gamma,\tau$) of the sinh-arcsinh normal distribution.  We train a fully connected neural network with five hidden layers of 100 units each using the {\tt ReLU} activation, a learning rate of 0.00005, and the {\tt Keras SGD} optimizer (see code for details). We train the model on 30,000 training samples, using early stopping based on the loss of 5,000 validation samples to optimize the number of training epochs. We then analyze an additional 5,000 samples as testing data. 

Figure~\ref{fig_clim_log} depicts the true $y$ label on the x-axis and the predicted distribution of $y$ on the y-axis, where the 80th percentile of the predicted distribution (i.e. 10th to 90th percentile of the distribution) is shown as a vertical blue bar. Purple and red dots depict the medians of the distributions, where purple denotes that the true $y$ falls within the 80th percentile range (``hit'') and red denotes that it does not (``miss''). Thus, for a perfect model, one would expect 80\% hits and 20\% misses. Instead, this neural network has 73\% hits and 27\% misses, suggesting that it tends to be slightly overconfident (i.e. the distribution is not wide enough). Either way, the network appears to do a relatively good job fitting the noisy data, and correctly identifies the inputs associated with positive $y$ labels as being noisier. It is worth noting that in the original paper, \cite{Mamalakis2021} required a much more complex network\footnote{In fact, they required a network with six hidden layers with 512, 256, 128, 63, 32, and 16 neurons, respectively!} to predict the non-noisy data with high accuracy.

\begin{figure}[!htb]
  \centering
  \includegraphics[width=4.0in]{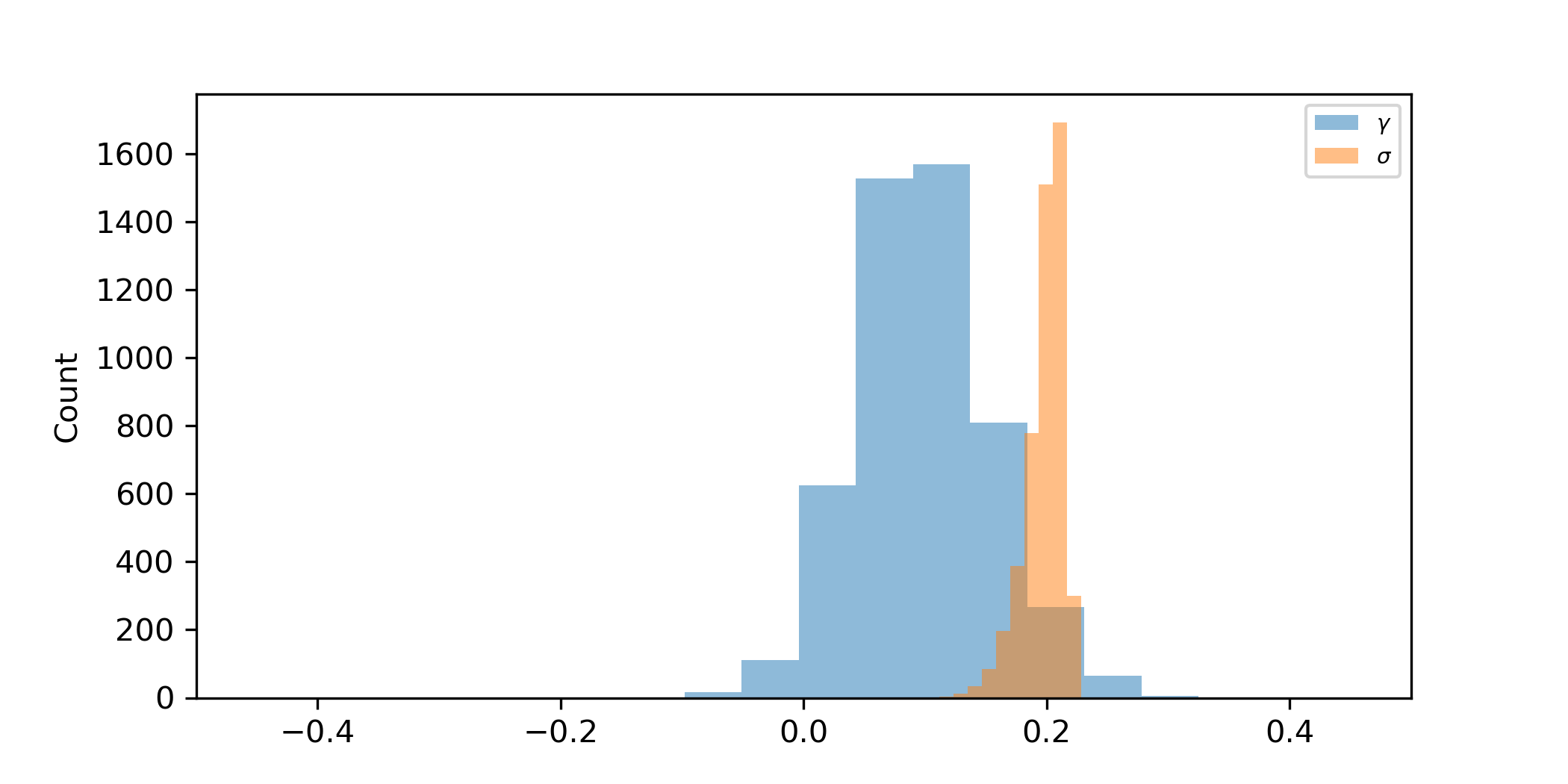}
  \caption{Predicted $\sigma$ and $\gamma$ for the synthetic climate data.}
  \label{fig_clim_log_hist} 
\end{figure}

It is difficult to visualize the skewness of the predicted conditional sinh-arcsinh normal distributions in Figure~\ref{fig_clim_log}, and so the predicted parameters $\sigma$ and $\gamma$ for the testing set are shown in Figure~\ref{fig_clim_log_hist}. From this we see that the network most often predicts a $\sigma$ of about 0.2 and a $\gamma$ (skewness) that is most often positive, reflective of the positively skewed noise we added to the data.

\begin{figure}[!htb]
  \centering
  \includegraphics[width=3.5in]{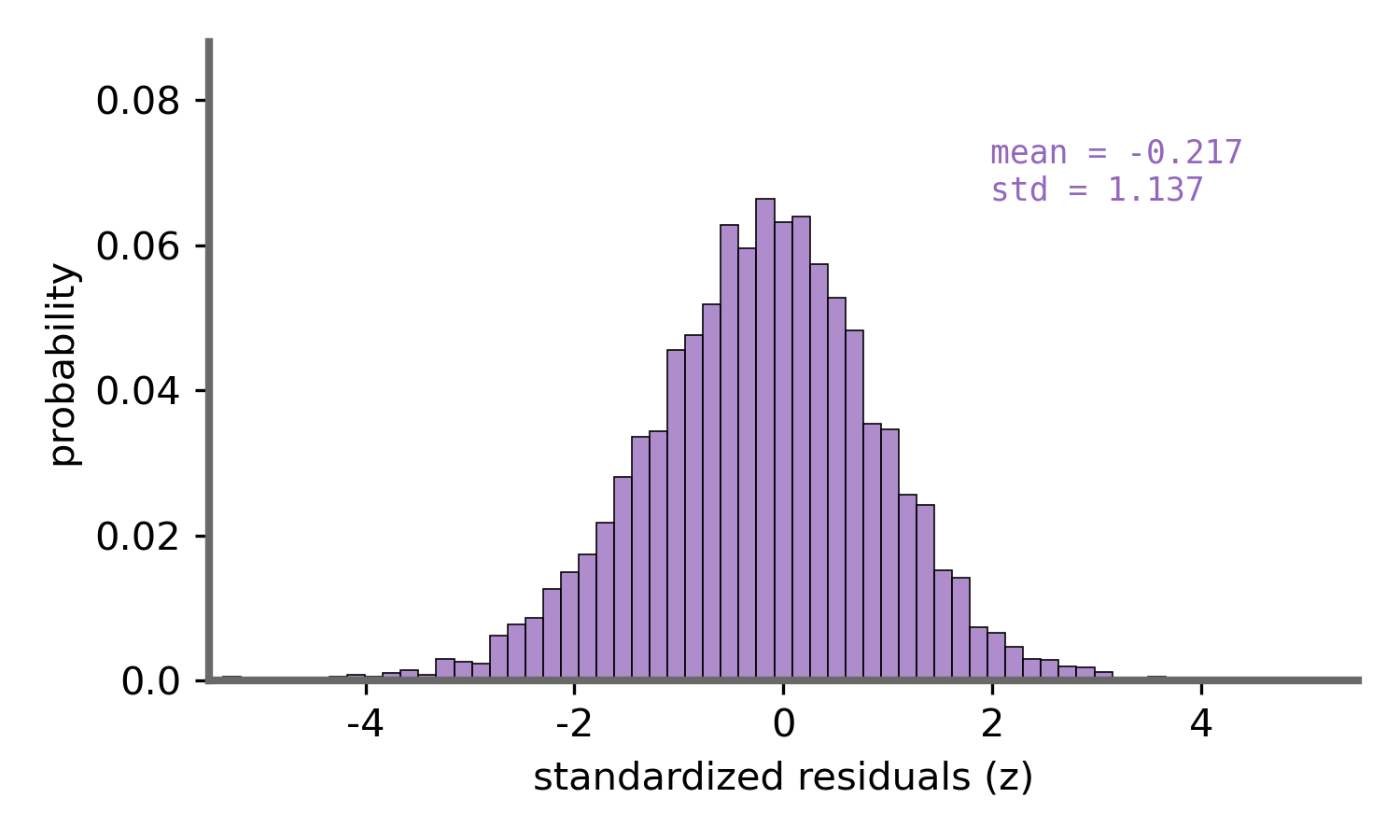}
  \caption{Histogram of standardized residuals of the testing set from the case of the synthetic climate data.}
  \label{fig_zscore_clim}
\end{figure}

As discussed in Section~\ref{S_diag}, there are many diagnostics worth running to assess whether the neural network's probabilistic predictions are of any use. Figure~\ref{fig_zscore_clim} shows the standardized residuals for the testing set of the synthetic climate. As in our previous example, the mean is close to $0$ and the standard deviation is close to $1$, suggesting that the network is generally producing meaningful results.

We have additionally evaluated the sign test as a check for how well the model predicts the middle of the conditional distribution. With a testing set of 5,000 samples, we find that $n_p = 2320$ and $n_n= 2680$. These values suggest that the network is preferentially slightly underestimating the median of the testing set.

\begin{figure}[!htb]
  \centering
  \includegraphics[width=4.0in]{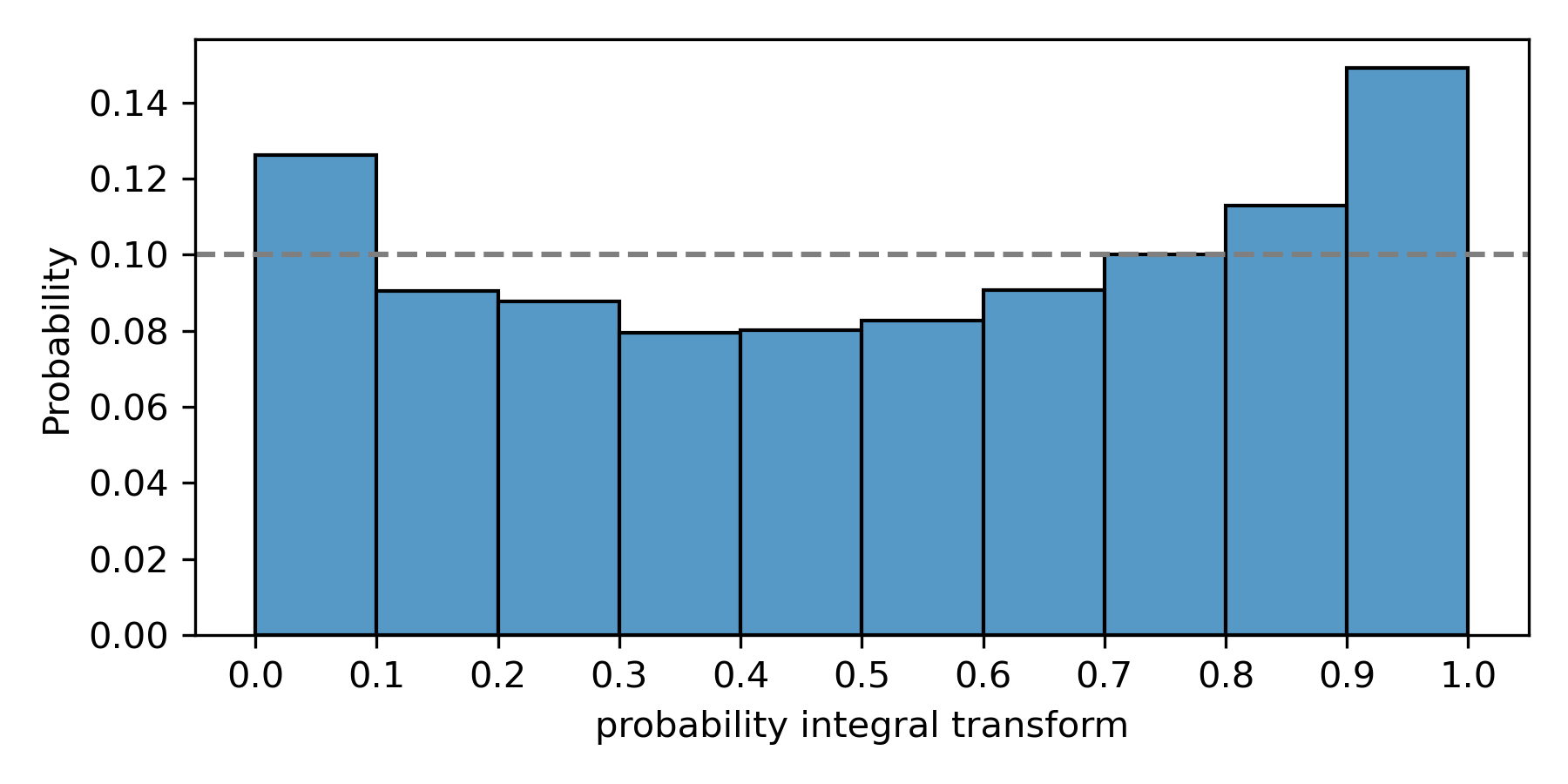}
  \caption{Probability integral transform (PIT) histogram of the testing set from the synthetic climate data. The horizontal dashed line denotes an equal probability of 0.1 across all bins.}
  \label{fig_pit_clim}
\end{figure}

In Figure~\ref{fig_pit_clim}, we show the PIT histogram for the synthetic climate data. This histogram shows more frequent PIT values at the two ends, suggesting that the network may tend to be over-confident in its predictions (i.e. on average, the conditional distributions are too narrow). The frequency of the other bins are generally similar and slightly below 10\%.  This blemish in the PIT histogram does not demand that we discard the model. Nonetheless, we must keep this overconfidence in mind as we apply the model, and we could consider alternate forms for the conditional distribution \cite{Scheuerer_2015}.

\section{Concluding Thoughts}
This simple method for incorporating uncertainty into neural network regression architectures has been previously presented as a standard approach in \cite{Duerr2020}, the {\tt Tensorflow Probability} manual, and many data science blogs. Even so, it appears relatively unknown in the geoscience community (a few exceptions include \cite{Foster2021}, \cite{BarnesBarnes2021} and \cite{Guillaumin2021}). The approach is simple, can be applied to most any network architecture, and can be adapted to different uncertainty distributions via the choice of the underlying distribution. Here, we explore the family of sinh-arcsinh normal distributions \cite{Jones2009} as a general set that can capture location, spread, and skewness in an intuitive manner. However, we note that other distributions (e.g. the lognormal distribution) may be better choices for specific applications.

It is worth contrasting the simple approach discussed here with that of Bayesian neural networks, which have become a go-to method for incorporating uncertainty into artificial neural networks (e.g. \cite{wilson2020case,Foster2021}). While incredibly powerful, the suite of Bayesian approaches is typically applicable when one has already defined a prior and wishes to update it with additional information. The simple approach described here is, instead, applicable when no such prior exists and one just wishes to parameterize the output and its uncertainty according to some previously defined family of distributions. The authors believe it will become a powerful, go-to method moving forward.

\section{Acknowledgements}
This work was funded, in part, by the NSF AI Institute for Research on Trustworthy AI in Weather, Climate, and Coastal Oceanography (AI2ES) under NSF grant ICER-2019758.

\clearpage

\bibliographystyle{hieeetr}
\bibliography{bibfile}
\clearpage

\appendix
\section{{\tt Python} Code to Generate the Synthetic Data}\label{A_code}
\subsection*{$x$ values}
\begin{lstlisting}[language=Python]
import numpy as np

N_SAMPLES = 25000
NP_SEED = 12345

np.random.seed(NP_SEED)
x = np.random.random(N_SAMPLES)
\end{lstlisting}

\subsection*{Linear, homoscedastic, normal (Figures~\ref{bg_01} and \ref{bg_02})}
\begin{lstlisting}[language=Python]
mu = 2 * x + 1
sigma = 0.75

y = np.random.normal(loc=mu, scale=sigma, size=N_SAMPLES)
\end{lstlisting}

\subsection*{Linear, homoscedastic, lognormal (Figure~\ref{bg_03})}
\begin{lstlisting}[language=Python]
mu = 2 * x + 1
sigma = 0.75

beta2 = np.log(1 + np.square(sigma / mu))
beta = np.sqrt(beta2)
alpha = np.log(mu) - beta2 / 2
               
y = np.random.lognormal(mean=alpha, sigma=beta, size=N_SAMPLES)
\end{lstlisting}

\subsection*{Linear, heteroscedastic, normal (Figure~\ref{bg_04})}
\begin{lstlisting}[language=Python]
mu = 2 * x + 1
sigma = 0.45 + 0.30 * np.cos(x * 3 * np.pi)

y = np.random.normal(loc=mu, scale=sigma, size=N_SAMPLES)
\end{lstlisting}

\subsection*{Nonlinear, heteroscedastic, normal (Figure~\ref{bg_05})}
\begin{lstlisting}[language=Python]
mu = 2 * x + 1 + 0.75 * np.sin(x * 3 * np.pi)
sigma = 0.45 + 0.30 * np.cos(x * 3 * np.pi)

y = np.random.normal(loc=mu, scale=sigma, size=N_SAMPLES)
\end{lstlisting}

\subsection*{Nonlinear, heteroscedastic, symmetric (Figure~\ref{fig_norm})}
\begin{lstlisting}[language=Python]
eps = (
    0.2 * np.random.normal(scale=1.0, size=N_SAMPLES)
    * 0.4 * np.cos(x * 1.75 * np.pi)
)

y = 2 * x + 1 + eps + 0.5 * np.sin(3 * x * np.pi)
\end{lstlisting}

\subsection*{Nonlinear, heteroscedastic, asymmetric (Figure~\ref{fig_lognorm})}
\begin{lstlisting}[language=Python]
eps = (
    0.2 * np.random.lognormal(mean=0.0, sigma=0.75, size=N_SAMPLES) 
    * 0.4 * np.cos(x * 1.75 * np.pi)
)

y = 2 * x + 1 + eps + 0.5 * np.sin(3 * x * np.pi)
\end{lstlisting}

\newpage
\section{Mapping to Enforce Parameter Constraints}\label{A_constraints}
\subsection{What and why?}
Consider the example given in Section~\ref{shash}. The  sinh-arcsinh normal distribution is defined by four parameters: $\mu$ (location), $\sigma$ (scale), $\gamma$ (skewness), and $\tau$ (tailweight). The parameters $\sigma$ (scale) and $\tau$ (tailweight) must be strictly positive. During training, our network does not know about these requirements. The output from our neural network is four real numbers: the second and fourth numbers are not necessarily positive.

To guarantee that $\sigma$ (scale) and $\tau$ (tailweight) are  strictly positive, we apply a mathematical trick called a {\it bijective map}. We reinterpret the second and fourth outputs (say $p$ and $q$) for each sample as $p = \log{(\sigma)}$ and $q = \log{(\tau)}$.  Then we compute $\sigma = \exp{(p)}$ and $\tau = \exp{(q)}$. Note that $-\infty < p < \infty$ so $0 < \sigma$, and $-\infty < q < \infty$ so $0 < \sigma$. A code fragment demonstrating this trick is given in Appendix~\ref{D_loss}.

Using {\it bijective maps}, we can enforce various constraints on the parameters. Three examples are given below. Note that the notations $x$ and $y$ in these examples are generic: they do not denote the components of observations.

\subsection{Strictly positive} \label{A_positive}
Example use: the standard deviation of a normal distribution must be positive. 
\begin{equation}
    y = \log{(x)}
\end{equation}
\begin{equation}
    x = \exp{(y)}
\end{equation}

\subsection{Between $0$ and $1$}
Example use: the scale of a log-logistic distribution must fall between $0$ and $1$ for the moments to be defined.

\begin{equation}
    y = \log{\left( \frac{1}{x} - 1 \right)}
\end{equation}
\begin{equation}
    x = \frac{1}{1 + e^y}
\end{equation}

\subsection{Between $a$ and $b$}
Example use: the mode of a von Mises distribution must be an angle (direction) between $0$ and $2\pi$.

\begin{equation}
    y = \log{\left( \frac{1}{\frac{x}{b-a} - a} - 1\right)}
\end{equation}
\begin{equation}
    x = \left( \frac{1}{1 + e^y} + a \right) \cdot (b - a)
\end{equation}
where $a < b$.

\clearpage
\section{Loss function example Tensorflow code}\label{D_loss}

\begin{lstlisting}[language=Python]
import tensorflow as tf
import tensorflow_probability as tfp


def RegressLossExpSigma(y_true, y_pred):
    mu   = y_pred[:, 0]
    std  = tf.math.exp(y_pred[:, 1])
    skew = y_pred[:, 2]
    tau  = tf.math.exp(y_pred[:, 3])

    cond_dist = tfp.distributions.SinhArcsinh(
        loc=mu, scale=std, skewness=skew, tailweight=tau
    )
    loss = -cond_dist.log_prob(y_true[:, 0])
    
    return tf.reduce_mean(loss, axis=-1)
\end{lstlisting}

In this code fragment, we are using the sinh-arcsinh normal distribution.  This distribution has four defining parameters.  As such, {\tt y\_pred} has four columns, one for each parameter of the local conditional distribution (see Appendix~\ref{C_network}).

Note that the extraction of {\tt std} and {\tt tau} from {\tt y\_pred} uses the trick described in Section~\ref{A_positive} of Appendix~\ref{A_constraints} to ensure that {\tt scale} and {\tt tailweight} are strictly positive.

\clearpage
\section{Network Architecture Example Tensorflow Code}\label{C_network}
\begin{lstlisting}[language=Python]
import tensorflow as tf
from tensorflow.keras.initializers import RandomNormal, Zeros

INPUT_SHAPE = 1
N_HIDDENS = 10
SEED = 99

inputs = tf.keras.Input(shape=INPUT_SHAPE)
x = inputs

# initialize a single hidden layer
x = tf.keras.layers.Dense(
    N_HIDDENS,
    activation="relu",
    use_bias=True,
    bias_initializer=RandomNormal(seed=SEED),
    kernel_initializer=RandomNormal(seed=SEED),
)(x)

# set final output units separately
mu_unit = tf.keras.layers.Dense(
    1,
    activation="linear",
    use_bias=True,
    bias_initializer=RandomNormal(seed=SEED),
    kernel_initializer=RandomNormal(seed=SEED),
)(x)

logsigma_unit = tf.keras.layers.Dense(
    1,
    activation="linear",
    use_bias=True,
    bias_initializer=Zeros(),
    kernel_initializer=Zeros(),
)(x)

skew_unit = tf.keras.layers.Dense(
    1,
    activation="linear",
    use_bias=True,
    bias_initializer=Zeros(),
    kernel_initializer=Zeros(),
)(x)

logtau_unit = tf.keras.layers.Dense(
    1,
    activation="linear",
    use_bias=True,
    bias_initializer=Zeros(),
    kernel_initializer=Zeros(),
)(x)

# final output layer
output_layer = tf.keras.layers.concatenate(
    [mu_unit, logsigma_unit, skew_unit, logtau_unit], axis=1
)

# finalize the model
model = tf.keras.models.Model(inputs=inputs, outputs=output_layer)
\end{lstlisting}

\clearpage
\section{Missing SHASH Helper Functions}\label{D_helper}
The {\tt Tensorflow Probability} documentation for the sinh-arcsinh normal distribution 
lists functions for mean, stddev, and variance, but they are not yet (v. 0.13.0)
implemented in the library. These following functions fill this gap.

\begin{lstlisting}[language=Python]
"""sinh-arcsinh normal distribution helper functions.

Functions
---------
mean(mu, sigma, gamma, tau)
    distribution mean.

median(mu, sigma, gamma, tau)
    distribution median.

stddev(mu, sigma, gamma, tau)
    distribution standard deviation.

variance(mu, sigma, gamma, tau)
    distribution variance.


Notes
-----
* The sinh-arcsinh normal distribution was defined in [1]. A more accessible
presentation is given in [2]. 

* The notation and formulation used in this code was taken from [3], page 143.
In the gamlss.dist/CRAN package the distribution is called SHASHo. 

* There is a typographical error in the presentation of the probability 
density function on page 143 of [3]. There is an extra "2" in the denomenator
preceeding the "sqrt{1 + z^2}" term.

References
----------
[1] Jones, M. C. & Pewsey, A., Sinh-arcsinh distributions,
Biometrika, Oxford University Press, 2009, 96, 761-780.
DOI: 10.1093/biomet/asp053.

[2] Jones, C. & Pewsey, A., The sinh-arcsinh normal distribution,
Significance, Wiley, 2019, 16, 6-7.
DOI: 10.1111/j.1740-9713.2019.01245.x.
https://rss.onlinelibrary.wiley.com/doi/10.1111/j.1740-9713.2019.01245.x

[3] Stasinopoulos, Mikis, et al. (2021), Distributions for Generalized 
Additive Models for Location Scale and Shape, CRAN Package.
https://cran.r-project.org/web/packages/gamlss.dist/gamlss.dist.pdf

"""
import numpy as np
import scipy
import tensorflow as tf

__author__ = "Randal J. Barnes and Elizabeth A. Barnes"
__date__ = "10 September 2021"


def _jones_pewsey_P(q):
    """P_q function from page 764 of [1].
    
    This is a module private helper function.  This function will not be 
    called externally.
    
    Arguments
    ---------
    q : float or double, array like
    
    Returns
    -------
    P_q : array like of same shape as q.
    
    Notes
    -----
    * The strange constant 0.25612... is "sqrt( sqrt(e) / (8*pi) )" computed 
    with a high-precision calculator.
    
    """
    return 0.25612601391340369863537463 * (
        scipy.special.kv((q + 1) / 2, 0.25) + scipy.special.kv((q - 1) / 2, 0.25)
    )


def mean(mu, sigma, gamma, tau):
    """The distribution mean.
    
    Arguments
    ---------
    mu : float or double (batch size x 1) Tensor 
        The location parameter. 
        
    sigma : float or double (batch size x 1) Tensor 
        The scale parameter. Must be strictly positive. Must be the same shape
        and dtype as mu.
        
    gamma : float or double (batch size x 1) Tensor 
        The skewness parameter. Must be the same shape and dtype as mu.
    
    tau : float or double (batch size x 1) Tensor 
        The tail-weight parameter. Must be strictly positive. Must be the same
        shape and dtype as mu.
    
    Returns
    -------
    x : float or double (batch size x 1) Tensor.
        The computed distribution mean values.
    
    """
    evX = tf.math.sinh(gamma / tau) * _jones_pewsey_P(1.0 / tau)
    return mu + sigma * evX


def median(mu, sigma, gamma, tau):
    """The distribution median.
    
    Arguments
    ---------
    mu : float or double (batch size x 1) Tensor 
        The location parameter. 
        
    sigma : float or double (batch size x 1) Tensor 
        The scale parameter. Must be strictly positive. Must be the same shape
        and dtype as mu.
        
    gamma : float or double (batch size x 1) Tensor 
        The skewness parameter. Must be the same shape and dtype as mu.
    
    tau : float or double (batch size x 1) Tensor 
        The tail-weight parameter. Must be strictly positive. Must be the same
        shape and dtype as mu.
    
    Returns
    -------
    x : float or double (batch size x 1) Tensor.
        The computed distribution mean values.

    Notes
    -----
    * This code uses the basic formula:
    
        E(a*X + b) = a*E(X) + b
        
    * The E(X) is computed using the moment equation given on page 764 of [1].
    
    """
    return mu + sigma * tf.math.sinh(gamma / tau)


def stddev(mu, sigma, gamma, tau):
    """The distribution standard deviation.
    
    Arguments
    ---------
    mu : float or double (batch size x 1) Tensor 
        The location parameter. 
        
    sigma : float or double (batch size x 1) Tensor 
        The scale parameter. Must be strictly positive. Must be the same shape
        and dtype as mu.
        
    gamma : float or double (batch size x 1) Tensor 
        The skewness parameter. Must be the same shape and dtype as mu.
    
    tau : float or double (batch size x 1) Tensor 
        The tail-weight parameter. Must be strictly positive. Must be the same
        shape and dtype as mu.
    
    Returns
    -------
    x : float or double (batch size x 1) Tensor.
        The computed distribution mean values.
    
    """
    return tf.math.sqrt(variance(mu, sigma, gamma, tau))


def variance(mu, sigma, gamma, tau):
    """The distribution variance.
    
    Arguments
    ---------
    mu : float or double (batch size x 1) Tensor 
        The location parameter. 
        
    sigma : float or double (batch size x 1) Tensor 
        The scale parameter. Must be strictly positive. Must be the same shape
        and dtype as mu.
        
    gamma : float or double (batch size x 1) Tensor 
        The skewness parameter. Must be the same shape and dtype as mu.
    
    tau : float or double (batch size x 1) Tensor 
        The tail-weight parameter. Must be strictly positive. Must be the same
        shape and dtype as mu.
    
    Returns
    -------
    x : float or double (batch size x 1) Tensor.
        The computed distribution mean values.
    
    Notes
    -----
    * This code uses two basic formulas:
    
        var(X) = E(X^2) - (E(X))^2
        var(a*X + b) = a^2 * var(X)

    * The E(X) and E(X^2) are computed using the moment equations given on
    page 764 of [1].
    
    """
    evX = tf.math.sinh(gamma / tau) * _jones_pewsey_P(1.0 / tau)
    evX2 = (tf.math.cosh(2 * gamma / tau) * _jones_pewsey_P(2.0 / tau) - 1.0) / 2
    
    return tf.math.square(sigma) * (evX2 - tf.math.square(evX))
\end{lstlisting}

\end{document}